\documentstyle[psfig]{jfm}
\newcommand{\be}{\begin{equation}}
\newcommand{\ee}{\end{equation}}
\newcommand{\lb}{\label}

\newcommand{\bh}{{\bf h}}
\newcommand{\bk}{{\bf k}}
\newcommand{\bp}{{\bf p}}
\newcommand{\bq}{{\bf q}}

\newcommand{\bu}{{\bf u}}
\newcommand{\bx}{{\bf x}}

\newcommand{\ou}{{\overline{u}}}
\newcommand{\ov}{{\overline{v}}}
\newcommand{\ow}{{\overline{w}}}
\newcommand{\bOm}{{\mbox{\boldmath $\Omega$}}}
\newcommand{\bom}{{\mbox{\boldmath $\omega$}}}
\newcommand{\bdelta}{{\mbox{\boldmath $\delta$}}}

\newcommand{\grad}{{\mbox{\boldmath $\nabla$}}}
\newcommand{\btimes}{{\mbox{\boldmath $\times$}}}
\newcommand{\bdot}{{\mbox{\boldmath $\cdot$}}}

\newcommand{\bnu}{{\bf \nu}}

\title[Resonant Interactions in Turbulence]
{Resonant Interactions in Rotating Homogeneous Three-dimensional Turbulence}

\author[Q. Chen, S. Chen, G. L. Eyink and D. D. Holm]%
{Q\ls I\ls A\ls O\ls N\ls I\ls N\ls G\ns C\ls H\ls E\ls N$^1$,
\ns S\ls H\ls I\ls Y\ls I\ns C\ls H\ls E\ls N$^{1,2,3}$, \ns
G\ls R\ls E\ls G\ls O\ls R\ls Y\ns L.\ns E\ls Y\ls I\ls N\ls
K$^{1,3}$\ns \and D\ls A\ls R\ls R\ls Y\ls L\ns D.\ns H\ls O\ls L\ls M$^{4,5}$}

\affiliation{
$^1$Department of Mechanical Engineering, The Johns Hopkins University,
Baltimore, MD 21218, USA\\[\affilskip]
$^2$CCSE and LTCS, Peking University, China\\[\affilskip]
$^3$Department of Applied Mathematics and Statistics, The Johns
Hopkins University,
Baltimore, MD 21218, USA\\[\affilskip]
$^4$Center for Nonlinear Studies and Theoretical Division, Los Alamos
National Laboratory, Los Alamos, NM 87545, USA\\[\affilskip]
$^5$Mathematics Department,
Imperial College London, SW7 2AZ, UK\\[\affilskip]
}
\date{?? and in revised form ??}

\begin{document}

\maketitle

\begin{abstract}
Direct numerical simulations of three-dimensional (3D) homogeneous
turbulence under rapid rigid rotation are conducted to examine the
predictions of resonant wave theory for both small Rossby number
and large Reynolds number. The theory predicts that ``slow modes''
of the velocity, with zero wavenumber parallel to the rotation
axis ($k_z=0$), will decouple from the remaining ``fast modes''
and solve an autonomous system of two-dimensional (2D)
Navier-Stokes equations for the horizontal velocity components,
normal to the rotation axis, and a 2D passive scalar equation for
the vertical velocity component, parallel to the rotation axis.
The Navier-Stokes equation for three-dimensional rotating
turbulence is solved in a $128^3$ mesh after being diagonalized
via ``helical decomposition'' into normal modes of the Coriolis
term. A force supplies constant energy input at intermediate
scales. To verify the theory, we set up parallel simulations for
the 2D Navier-Stokes equation and 2D passive scalar equation to
compare them with the slow-mode dynamics of the 3D rotating
turbulence. The simulation results reveal that there is a clear
inverse energy cascade to the large scales, as predicted by 2D
Navier-Stokes equations for resonant interactions of slow modes.
As the rotation rate increases, the vertically-averaged horizontal
velocity field from 3D Navier-Stokes converges to the velocity
field from 2D Navier-Stokes, as measured by the energy in their
difference field. Likewise, the vertically-averaged vertical
velocity from 3D Navier-Stokes converges to a solution of the 2D
passive scalar equation. The slow-mode energy spectrum approaches
$k_h^{-5/3},$ where $k_h$ is the horizontal wavenumber, and energy
flux becomes closer to constant, as in 2D, the greater the
rotation rate. Furthermore, the energy flux directly into small
wave numbers in the $k_z=0$ plane from non-resonant interactions
decreases, while fast-mode energy concentrates closer to that
plane. The simulations are consistent with an increasingly
dominant role of resonant triads for more rapid rotation.

\end{abstract}

\section{Introduction}

Large-scale flows in oceans and the atmosphere are greatly
affected by the earth's rotation and are known to be
quasi-two-dimensional. Rotation also plays an important role in
many engineering flows, e.g. high-Reynolds turbulent flows in
turbo-machinery and rotating channels. When a fluid is under
rotation, the Coriolis force is introduced into the momentum
equation and competes with the nonlinear force. A dimensionless
number, the Rossby number $Ro,$ may be defined as the ratio of the
magnitude of the inertial term to the Coriolis force. In a rapidly
rotating fluid, a mathematical limit $Ro \rightarrow 0$ is taken
of the three-dimensional Navier-Stokes equations. There is a
general expectation that the fluid should become two-dimensional
in this limit. For example, the classic Taylor-Proudman theorem
implies that the Coriolis force will align vortex tubes parallel
to the rotation axis in steady-state, rapidly-rotating flows
(\cite{Green68}).

To take into account the effect of rapid rotation on dynamical
evolution a {\it resonant wave theory} has been applied
(\cite{Green68,Green69,Waleffe93}). Similar theories were first
developed for gravity waves in geophysical fluid flows
(\cite{Phillips60}) and have since been widely invoked elsewhere.
For excellent reviews, see \cite{Phillips81} and \cite{Craik85}.
According to this approach, the fluid velocity may be regarded in
the limit of small Rossby number as a superposition of inertial
waves with a large characteristic frequency which are modulated on
a longer, slow time-scale. An averaged equation is derived from
weakly nonlinear theory for the slow-time motion. This equation
explains enhanced energy transfer from small scales to large
scales by the resonant triadic interaction of inertial waves
(\cite{Green68,Green69,Waleffe93}). Using an ``instability
hypothesis'', \cite{Waleffe93} argued further that resonant
triadic interactions should drive the flow to become
quasi-two-dimensional. More recently, this type of resonant wave
theory has been put on a sounder mathematical footing.
\cite{EM96,EM98} and \cite{ME98} have derived similar ``averaged
equations'' in a rigorous asymptotic limit over a fixed time
interval for a general class of geophysical fluids problems with
fast wave dynamics. For the particular case of rotating
incompressible fluids, it has been shown that the averaged
equations contain as a subset the 2D Navier-Stokes equations for
the vertically-averaged velocity fields (\cite{MZ96,BMN96}).

However, in {\it turbulent} flow under rapid rotation not only is
the Rossby number $Ro$ small, but also the Reynolds number $Re$ is
large. As a consequence, eddy motions are excited on a wide range
of length scales. At any given wavenumber $k$ there are thus at
least two distinct time scales, the rotation time scale
$\tau_\Omega \sim \frac{1}{2\Omega}$ and the nonlinear time scale
$\tau_{non}(k) \sim (k^3 E(k))^{-1/2}$ (where $E(k)$ is the energy
spectrum). The validity of the wave resonance theories depend upon
$\tau_\Omega$ being shorter than all other time scales in the
problem. However, for very large Reynolds numbers, there will be a
range of high wavenumbers $k$ where instead $\tau_{non}(k) \ll
\tau_\Omega.$ Thus, the resonant wave theory is likely to be valid
for turbulent flows only very non-uniformly in wavenumber, if at
all. The existing mathematical derivations of the theory
(\cite{EM96,EM98,ME98,MZ96,BMN96}) are only usefully valid for low
Reynolds numbers. Indeed, error fields for the resonant wave
approximation are estimated in these proofs by ``Sobolev norms''
that get most of their contribution from high-wavenumbers. The
errors are therefore shown, by present theorems, only to have
upper bounds that grow rapidly with the Reynolds number $Re.$
While the error bounds also decrease with Rossby number $Ro,$
arguments based on comparing time-scales, like those above,
suggest that resonant triads will be selected latest at the
highest wavenumbers. Thus, very low Rossby numbers should be
required in the existing proofs to guarantee that errors in the
wave resonance theory are small for fully-developed turbulent
flow.

The present evidence from simulations and experiments is also
mixed as to the validity of the theory. Simulations of decaying
rotating turbulence by \cite{BFR85} showed a tendency for the
length-scales along the axis of rotation to grow as rotation rate
increases. \cite{Bartello94} observed two-dimensional vortices
emerging from the three-dimensional flow. \cite{Hoss94} showed
that the turbulent flow reduced to an approximate two-dimensional
state and the energy cascaded to longer length scales in a $128^3$
forced simulation. These results are roughly in accord with the
wave resonance theory. The first numerical work to study
explicitly the relation of flow two-dimensionalization and
resonant triadic interactions was \cite{LW99}. They observed a
strong two-dimensionalization and a clear inverse energy cascade.
However, they suggested that non-resonant interactions might still
play an important role at rotation rates achieved in their
simulations. So far the detailed predictions of 2D turbulence
theory have not been verified numerically in rotating 3D
turbulence and the role of resonant wave interactions has remained
unclear for Reynolds number $Re\gg 1$. Thus, it is the purpose of
this paper to investigate further the limit of rapidly rotating 3D
turbulence, where resonant interactions should dominate and 2D
turbulence be achieved.

The remainder of this paper is organized as follows. In section 2,
we review the resonant wave theory of rapidly rotating fluids,
including the rigorous mathematical results. Particular attention
will be paid to a ``Dynamic Taylor-Proudman theorem'' that
predicts the time-evolution of the vertically-averaged or 2D
fields. In Appendix A, we give an elementary derivation of this
result based upon the ``averaged equation'' of the resonant wave
theory. In section 3, we discuss the numerical schemes and present
our simulation results to investigate the mechanism of
two-dimensionalization in rotating turbulence and to study the
role of resonant triadic interactions. Finally, our conclusions
are presented in section 4.

\section{Resonant Wave Theory}

Rapidly rotating fluids are a multi-time scale problem. In a
rotating frame of reference, the Navier-Stokes equation reads (see
\cite{Green68})
 \be \partial_t \bu + 2\bOm\btimes\bu =
  -\grad P/\rho + \nu \grad^2 \bu
- \bom \btimes \bu  \lb{N-S}. \ee
Here, rotation vector $\bOm=\Omega\widehat{{\bf z}}$.
$\bu(\bf{x})$ is the velocity field, $\bom=\grad \btimes \bu$ is
the vorticity field, $\rho$ is the density, $\nu$ is the kinematic
viscosity and $P$ is the pressure in an inertial frame modified by
a centrifugal term: $P=P_0+ \frac{1}{2}\rho\|\bOm\btimes{\bf
x}\|^2$. If $L$ and $U$ are characteristic length and velocity
scales, then the above equation is non-dimensionalized as
\be
\partial_{\tilde{t}} \tilde{\bu} + \frac{2}{Ro}
\widehat{\bf{z}} \btimes \tilde{\bu}
 = -\tilde{\grad} \tilde{P} + \frac{1}{Re} \,
 \tilde{\grad}^2 \tilde{\bu}
- \tilde{\bom} \btimes \tilde{\bu}  \lb{1_Dimen-N-S}. \ee
\noindent Here, Rossby number $Ro$ is defined as
\be Ro = \frac{U}{\Omega L},\ee
\noindent and Reynolds number $Re$ as
\be Re = \frac{UL}{\nu}. \ee
\noindent For simplicity, we get rid of the symbol tilde from now
on.

The first step in developing the resonant wave theory is to expand
the governing fields into ``normal modes'', i.e. the eigenmodes of
the linearized equation. In the case of rapid rotation this
corresponds to an expansion of the velocity field into {\it
helical modes}:
\be \bu(\bx,t) =
\sum_\bk\sum_{s=\pm}a_s(\bk,t)\bh_s(\bk)e^{i\bk\bdot\bx}
\lb{v-Hdecomp} \ee
where $\bh_{\pm}(\bk)$ are defined as orthogonal eigenmodes of the
curl operator, satisfying $i\bk\btimes\bh_s = s|\bk|\bh_s$ with
$s=\pm 1$ (\cite{Green68,Waleffe92}). In this basis, the Coriolis
term is diagonalized and (\ref{1_Dimen-N-S}) can be written as
 \be (\partial_t
- i \frac{1}{Ro} \omega_{s_k} + \frac{1}{Re} \,\, k^2)
a_{s_k}=\frac{1}{2} \sum_{\bk+\bp+\bq=0} \sum_{s_p,s_q}
C^{s_k,s_p,s_q}_{\bk\bp\bq} a^*_{s_p} a^*_{s_q}.\lb{heli-N-S} \ee
As a convenient shorthand, we abbreviate the pair $(\bk,s)$ as
$s_k$. The frequency $\omega_{s_k}=2s(\widehat{\bf{z}} \cdot
\bk)/k = 2 s k_z/k=2 s \cos\theta_k $, with $\theta_k$ the angle
between $\bOm$ and wavenumber vector $\bk$. The zero frequency
modes or so-called {\it slow modes}, with $\omega_{s_k}=0,$ are
precisely those with $k_z=0$ and thus coincide with 2D modes
having no variation along the rotation axis. The remaining modes
with $k_z \ne 0$ are fully 3D modes and, since they have nonzero
frequency, are called {\it fast modes}. The slow modes can also be
obtained by vertically averaging the 3D velocities $\overline{{\bf
u}}^{3D}(x,y)=\frac{1}{H}\int_0^{H} {\bf u}(x,y,z) dz,$ with $H$
the vertical height of the domain.

For $Ro \rightarrow 0$, the solution of Eq.(\ref{heli-N-S}) is
assumed in the resonant interaction theory to evolve on two
distinct time-scales, the slow time $t$ and the fast time scale
$\tau=\frac{t}{\epsilon}$ for $\epsilon=Ro.$ A standard
multiple-scale asymptotic expansion is then made with
$\partial_t\rightarrow\partial_t +\frac{1}{Ro}\partial_\tau.$ To
leading order, the solution satisfies a multiple-scale Ansatz $
a_{s_k}(t,\tau)= A_{s_k}(t) e^{i \omega_{s_k} \tau}$, consisting
of inertial waves with rapid oscillations on the fast time scale
and amplitude $A_{s_k}(t)$ depending on the slow time $t$. This
slow time-dependence is determined to eliminate secular terms
growing like $\tau$, yielding the ``averaged equation'':
\be (\partial_t + \frac{1}{Re} \,\, k^2) A_{s_k}=\frac{1}{2}
\sum^{\omega_{s_p}+\omega_{s_q}+\omega_{s_k}=0}_{\bk+\bp+\bq=0}
\sum_{s_p,s_q} C^{s_k,s_p,s_q}_{\bk\bp\bq} A^*_{s_p} A^*_{s_q},
\lb{averaged} \ee
valid for a slow time $t=O(1)$. Only ``resonant triads''
satisfying
\be \omega_{s_p}+\omega_{s_q}+\omega_{s_k} =0 \lb{resonance} \ee
still remain in this equation. See \cite{Green68,Waleffe93} for
more details.

There are three classes of resonant triads: ``fast-slow-fast'' and
``slow-slow-slow'' and ``fast-fast-fast''. We follow the
convention that the first wavenumber in the triad is the one
appearing on the left hand  of equation (\ref{averaged}), and is
thus the mode which undergoes evolution due to interaction of the
other two modes.  For example, a ``fast-slow-fast'' triad
gives a contribution to the evolution of a fast mode due to the
interaction of another fast mode and a slow mode. In such a
resonant triad, the slow mode acts simply as a catalyst for energy
exchange between the two fast modes and its own energy is
unchanged (\cite{Green69, Waleffe93}). Formally speaking, there
are resonant ``slow-fast-fast'' triads, but they have zero
coupling coefficient. This result is known to hold more generally
in the theory of resonant fluid interactions, not only for simple
rotation but also for $\beta$-plane flows
(\cite{Longuet-HigginsGill67}), stratified flows
(\cite{Phillips68,LeLongRiley91}), rotating stratified flows
(\cite{Bartello95}), and rotating shallow-water flows
(\cite{Warn86}). As a consequence, the slow 2D modes in the limit
of rapid rotation evolve under their own autonomous dynamics. This
consists of all the ``slow-slow-slow'' triadic interactions, each
of which is resonant. The averaged equation for the autonomous 2D
modes splits into two parts, as shown by \cite{EM96,BMN96}. As
$Ro\rightarrow 0,$ the vertically-averaged horizontal velocity
$\overline{\bu}_H^{3D}=(\ou^{3D},\ov^{3D})$ satisfies the 2D N-S
equation:
 \be \partial_t \overline{\bu}_H^{3D} + (\overline{\bu}_H^{3D}
 \bdot
   \grad)\overline{\bu}_H^{3D} =
  -\grad P_H/\rho + \nu \grad^2 \overline{\bu}_H^{3D}, \lb{2D_N-S} \ee
while the vertically-averaged
vertical velocity $\ow^{3D}$ satisfies the 2D passive scalar
equation:
\be
\partial_t \ow^{3D} + (\overline{\bu}_H^{3D}\bdot
   \grad)\ow^{3D} = \nu \grad^2 \ow^{3D}. \lb{2D_passive} \ee
We give an elementary (but nonrigorous) derivation of these
results in the Appendix. Notice that it is not implied by this
result that a flow under rapid rotation will become
two-dimensional, but it does mean that the dynamics will contain
an independent two-dimensional subdynamics. In this respect, the
result resembles the classic Taylor-Proudman theorem for steady
flows (\cite{Green68}), so that it can be termed the ``Dynamic
Taylor-Proudman Theorem.'' Precisely, the statement is that the
``slow-slow-slow'' triadic interactions yield the 2D-3C
Navier-Stokes equations. 2D-3C means two variables ($x$,$y$) but
three components $(\ou^{3D},\ov^{3D},\ow^{3D})$. Continuing with
the accounting of resonant triads, the final set with all fast
modes are called {\it fast-fast-fast}. The averaged equation
(\ref{averaged}) for the evolution of fully 3D modes contains
interactions of both the ``fast-slow-fast'' and ``fast-fast-fast''
types. While these cannot transfer energy directly to slow 2D
modes, \cite{Waleffe93} has argued that ``fast-fast-fast''
resonances do play an important role in flow
two-dimensionalization. As a consequence of an ``instability
hypothesis'', he has shown that energy in the fast 3D modes is
transferred by this set of resonances preferentially to slower
modes, with smaller (but not zero) vertical wavenumber $k_z.$

The multiple-scale Ansatz and the averaged equation
(\ref{averaged}) have been rigorously proved by \cite{EM96} and
\cite{ME98} for a general fluid dynamical equation with fast wave
dynamics, which includes rapidly-rotating fluids as a special
case. The precise statement of their result is that there exists
some finite time $T>0$ and an exponent $p>1+d/2$ (with $d$ space
dimension) such that, for all $0<t<T,$
\be \sum_{\bk,s_k} k^{2p}
\left|a_{s_k}^{Ro}(t)-A_{s_k}(t)e^{i\omega_{s_k}t/Ro}\right|^2=
o(1) \lb{limit} \ee
as $Ro\rightarrow 0$, where $a_{s_k}^{Ro}(t)$ is the solution of
the full equation (\ref{heli-N-S}) for given Rossby number $Ro,$
and $A_{s_k}(t)$ is the solution of the averaged equation
(\ref{averaged}). Thus the error in the multiple-scale
approximation goes to zero in the Sobolev-norm sense of
(\ref{limit}) as $Ro\rightarrow 0.$ This result can be stated in
another way, in terms of the error field
\be \bdelta^{Ro}(\bx,t)\equiv \bu^{Ro}(\bx,t)-{{\bf U}}(\bx;
t,t/Ro), \lb{error} \ee
where $\bu^{Ro}(\bx,t)$ is the solution of the rotating
Navier-Stokes equation (\ref{1_Dimen-N-S}) and ${\bf U}(\bx;
t,\tau)$ is the multiple-scale Ansatz written in physical space.
Define $E_\delta^{Ro}(k,t)$ as the wavenumber spectrum of the
field $\bdelta^{Ro}(\bx,t),$ or the {\it error energy spectrum}
(\cite{Kr70,Leith72}). Then (\ref{limit}) is equivalent to the
statement that, as $Ro\rightarrow 0,$
\be \int_0^\infty dk\,\,k^{2p}E_\delta^{Ro}(k,t)=o(1), \lb{limit2}
\ee
for all $0<t<T$ and some $p>1+d/2$. Thus, the theorem guarantees
that some moment of the error spectrum goes to zero, at least over
a finite time-interval, as $Ro\rightarrow 0.$

The multiple-scale argument applies to any rotating 3D flow,
whether laminar or turbulent. However, the error bounds in
(\ref{limit}) and (\ref{limit2}) assume that, for $d=3,$ energy
spectra decay faster than $k^{-6}$ at high wavenumbers $k,$ much
steeper than turbulent spectra in the inertial range. Of course,
when the Reynolds number $Re$ is large but finite, then the
spectra shall eventually decay exponentially at large enough $k,$
in the far dissipation range. However, because of the long
inertial range, the Sobolev norms in (\ref{limit}) and
(\ref{limit2}) will become strongly Reynolds-number dependent,
expected to grow as some power $(Re)^{\xi_p}.$ Thus, the norms
will not be small at high Reynolds number, except when the Rossby
number is extremely low. Indeed, note that the spectral moments in
(\ref{limit2}) will get most of their contribution from near the
Kolmogorov dissipation wavenumber $k_d$ in a turbulent flow. That
is the last wavenumber range where resonant triads will be
selected as $Ro\rightarrow 0,$ because the eddy turnover time
$\tau_{non}(k_d)$ in the dissipation range is the shortest in the
entire flow. Therefore, the present theorems effectively say
nothing about the validity of the resonant wave theory at high
$Re$, for realistic values of the Rossby number. Since previous
numerical simulations of rotating turbulence by
\cite{BFR85,Bartello94,Mansour92,Hoss94,LW99} have also not
verified the predictions of (\ref{averaged}), its status for
turbulent flow has remained unclear.

\section{Numerical Simulations and Analysis}

To address this issue, we have simulated the rotating N-S equation
Eq.(\ref{1_Dimen-N-S}) in a $128^3$ periodic box with a forcing
\be f_i({\bf k},t) =\epsilon_{\bk,i}/\hat{u}_i(\bf k, t)^*.\ee
Here, $\hat{u}_i(\bf k, t)^*$ is the conjugate of Fourier
component $\hat{u}_i(\bf k, t)$. This force is specially chosen so
that the energy input rates $\epsilon_{\bk,i}$ are all fixed
(\cite{KerrSigg78}), and we choose the latter to be constant in a
narrow band $22<k_f<24$ and zero elsewhere. Thus, the total input
power $\epsilon = \sum_{\bk,i} \epsilon_{\bk,i}$ is constant, as
well as the separate energy inputs into slow modes and fast modes.
This forcing guarantees also that the energy input is the same for
all rotation rates. The normal viscosity term $\nu \grad^2 \bu$ is
replaced by a hyperviscosity term $(-1)^{p+1}\nu_p (\grad)^{2p}
\bu$ with $p=8$, to extend the inertial range. The Coriolis force
and viscosity term are integrated exactly using a slaved,
2nd-order Adams-Bashforth scheme. At each time step, the velocity
in Fourier space $\hat \bu(\bk,t)$ is decomposed into the two
helical modes and these two modes are evolved according to
Eq.(\ref{heli-N-S}). The nonlinear term is calculated using the
usual pseudospectral method and then projected onto the two
helical modes (\cite{LW99}). The macro-Rossby number $Ro^L =
(\epsilon k^2_f)^{1/3}/\Omega$ ranges from 0.066 to 0.0021. Rossby
numbers from the various experimental and numerical studies are
collected in Table 1. In the table {\it micro-} Rossby number
$Ro^\omega=\omega'/(2\Omega)$ and $\omega'$ is {\it r.m.s}
vorticity. Among the direct numerical simulations of forced
rotating turbulence, \cite{YZ98} were interested in the dynamics
in the range $k>k_f$, while we and others (\cite{LW99},
\cite{Hoss94}) focus on the dynamics in the inverse energy cascade
range $k<k_f$. We consider a transient flow state, in which
rotation is begun after a statistically steady state is reached
without rotation. The results shown below, if without
specification, were obtained after taking an ensemble average over
eight realizations started from distinct initial conditions. This
simulation is patterned after a $128^2$ simulation of 2D N-S by
\cite{SY93}. They found that an inertial-range energy spectrum
$k^{-5/3}$ is established by the 2D inverse cascade process and
energy flux in the inverse cascade range is a negative constant,
at least before energy begins to accumulate at the largest scales.
To compare the results of our 3D rotating simulation with its
hypothetical limit, described by the 2D N-S and passive scalar
equations, we carry out simultaneous calculations with
Eq.(\ref{2D_N-S}) and Eq.(\ref{2D_passive}) in which the initial
conditions are the vertically-averaged initial conditions of the
rotating flow and the force is the vertically-averaged 3D force.
In such a setup the simulation time is also the slow time. With
these simulations we systematically check the main predictions of
the averaged equation Eq.(\ref{averaged}) for the resonant
interactions.

\smallskip
\begin{table}
\begin{center}
\begin{tabular}{lccc}
Experiment/Numerical Simulations   & Rossby Number  & Forcing Scale\\
$Traugott(1958) \,\ (decay)$ & $Ro^\omega=1.65$ & $None$\\
$Wigeland \,\ $\&$ \,\ Nagib(1978) \,\ (decay)$ & $Ro^\omega=0.4 \sim 16$ & $None$\\
$Bardina \,\ et \,\ al.(1985) \,\ (decay)$ & $Ro^\omega=0.3 \sim 6.3$ & $None$ \\
$Jacquin \,\ et \,\ al.(1990) \,\ (decay)$ & $Ro^L=0.2 \sim 12$ & $None$ \\
$Bartello \,\ et \,\ al.(1994)\,\ (decay)$ & $Ro^\omega=0.01 \sim 100$ & $None$ \\
$Hossain(1994) \,\ (forced)$ & $Ro^L=0.1$ & $11 \leq k^2_f \leq 13$\\
$Yeung \,\ $\&$ \,\ Zhou(1998) \,\ (forced)$ & $Ro^L=0.00064 \sim 0.0195$ & $k_f \leq 2$\\
$Smith \,\ $\&$ \,\ Waleffe(1999) \,\ (forced)$ & $Ro^L=0.17,0.35$ & $k_f =24$\\
$ Our \,\ 128^3 DNS(2003)\,\ (forced)$ & $Ro^L=0.0021 \sim 0.066$ & $22 \leq k_f \leq 24$\\
\end{tabular}
\caption{Different Rossby Numbers from Experimental and Numerical
Studies}
\end{center}
\end{table}
\smallskip

As Rossby number asymptotically approaches zero, the resonant
triadic interactions represented in the ``averaged equation''
should become more and more dominant over the non-resonant ones.
In Table 2, all resonant and non-resonant triadic interactions and
their characteristics in the rapid-rotation limit are listed. In
this section, we examine carefully the validity of the resonant
wave theory as we decrease the Rossby number. To be more specific,
slow-slow-slow triadic interactions, slow-fast-fast triadic
interactions and fast-fast-fast triadic interactions are studied,
respectively. We shall not study the catalytic ``fast-slow-fast''
resonant triads, because these can transfer energy only between
fast modes with the same wavevector magnitude $k$ and the same
value of $\cos\theta$ (\cite{Waleffe93}). Hence, they can play no
direct role in transferring energy between scales or in
two-dimensionalization of the flow. Their plausible effect is
simply to isotropize the fast mode energy distribution in the
horizontal wavenumber plane.

In Fig.~1 are plotted 3D energy spectra for different Rossby
numbers and also for the parallel 2D run. In the plot,
$k_h=(k_x^2+k_y^2)^{1/2}$ is horizontal wavenumber magnitude. When
rotation is ``turned on'', energy is transferred to the large
scales as shown in Fig.~1 (a) and (c), consistent with the
previous observations (\cite{Hoss94} and \cite{LW99}). Notice that
large-scale energy grows faster at $Ro=0.066$ than at $Ro=0.0021,$
when the energy input is the same. For Rossby number $Ro=0.17$,
\cite{LW99} observed a rapid energy transfer to low wavenumbers
similar to ours at $Ro=0.066$, which they interpreted as due to
fast, non-resonant interactions of inertial waves. For both of our
Rossby numbers, the flow tends to two-dimensionalize. In
particular, Fig.~1 (b) and (d) show that slow-mode energy
$E(k_h,k_z=0)$, energy from vertically-averaged horizontal
velocity $E_{uv}(k_h,k_z=0)$ and total energy $E(k)$ all collapse
together at large scales. This does not mean, however, that the
fluid dynamics is that of 2D-NS. At $Ro=0.066,$ these three
spectra are still far from the spectrum $E_{2D}(k_h)$ of the 2D-NS
solution. It is only at $Ro=0.0021$ that the spectra of $k_z=0$
nodes begin to approach the 2D-spectrum $E_{2D}(k_h)$ suggesting
that the resonant wave theory is becoming more valid for the lower
Rossby number.

\smallskip
\begin{table}
\begin{center}
\begin{tabular}{lccc}
Type of triad & Resonant/Non-resonant & Characteristics
as $Ro \rightarrow 0$ \\
$slow-slow-slow$ & $ R $ & $2D+3C$ \\
$slow-fast-fast$ & $ N $ & $vanishing$ \\
$fast-slow-fast$ & $ R + N $ & $catalytic$\\
$fast-fast-fast$ & $ R + N $ & $quasi-2D$\\
\end{tabular}
\caption{Triadic Interactions in Three-dimensional Rotating
Turbulence}
\end{center}
\end{table}
\smallskip

\smallskip
\centerline{\psfig{file=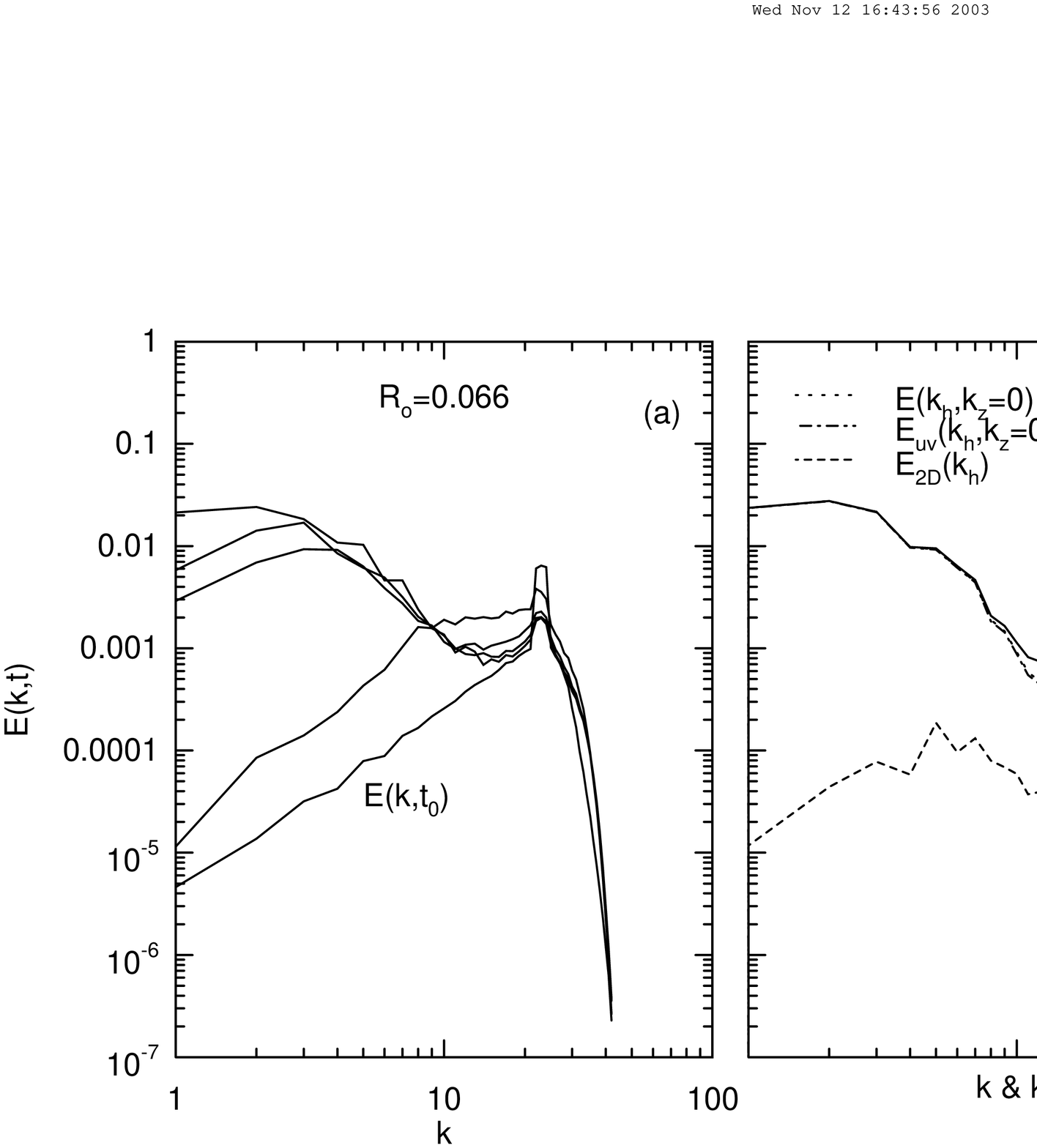,width=300pt,height=160pt}}
\centerline{\psfig{file=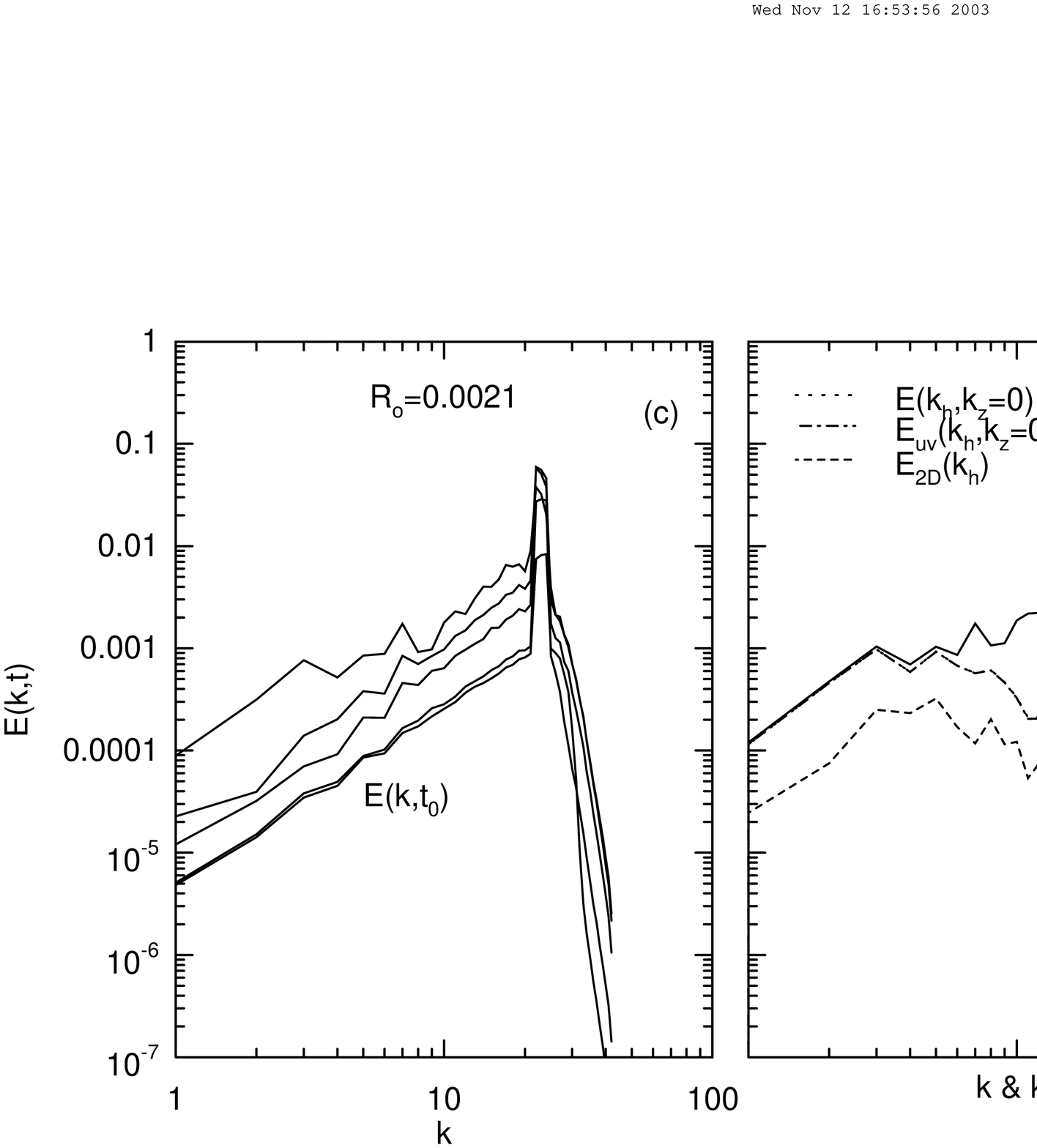,width=300pt,height=160pt}}
\noindent
 {\small FIGURE 1.Time evolution of energy spectra (upward) for
(a) $Ro=0.066$ and (c) $Ro=0.0021$ at $t=0$, $t=172\tau_0$,
$t=344\tau_0$, $t=430\tau_0$ and $t=600\tau_0$; (b) for
$Ro=0.066$, large-scale motion is strongly excited and
two-dimensional over a broad range of low-wavenumbers at
$t=600\tau_0$ (from one realization) and (d) for $Ro=0.0021$,
large-scale motion is weaker and two-dimensional over a smaller
range at $t=600\tau_0$ (from one realization). $E(k,t_0)$ is the
initial energy spectrum.}
\bigskip

To investigate how energy is transferred to the large scales, we
calculate the energy transfer functions $
T(k_h,k_z)=\sum_{S(k_h),I(k_z)} Re[\hat{\bu}^* \bdot \hat{\bu}
\times \hat{\bom}]$ at various rotation rates, where $S(k_h)$
denotes a circular shell of horizontal wavenumbers with central
radius $k_h$ and $I(k_z)$ is an interval of vertical wavenumbers
with midpoint $k_z.$ These energy transfer functions are plotted
in Figs.~2--4, normalized by their largest absolute values.
Initially before rotation is added, most of the energy transfer
activity happens in the vicinity of the forcing scale $k_f,$ where
 $T(k_h,k_z)$ is negative (see Fig.~2 (a) and Fig.~3 (a)).
 When the system rotates slowly, at the largest simulated Rossby number
$Ro=0.066$, energy is quickly carried away from the forcing scales
to the $k_z=0$ plane and further towards smaller $k_h$ (see
Fig.~2). There is a general tendency, at all Rossby numbers, for
the largest transfers at later times to be into the slow modes at
$k_z=0.$ However, when the system rotates faster, at $Ro=0.00825$,
there is less energy transfer to the large scales at $t=450
\tau_0$ (see Fig.~3). When the system rotates thirty times faster,
at $Ro=0.0021$ (see Fig.~4), most of the energy transfer is again
concentrated near the $k_z=0$ plane at $t=450 \tau_0$, but hardly
any transfer has developed to smaller $k_h$.

\bigskip
\centerline{\psfig{file=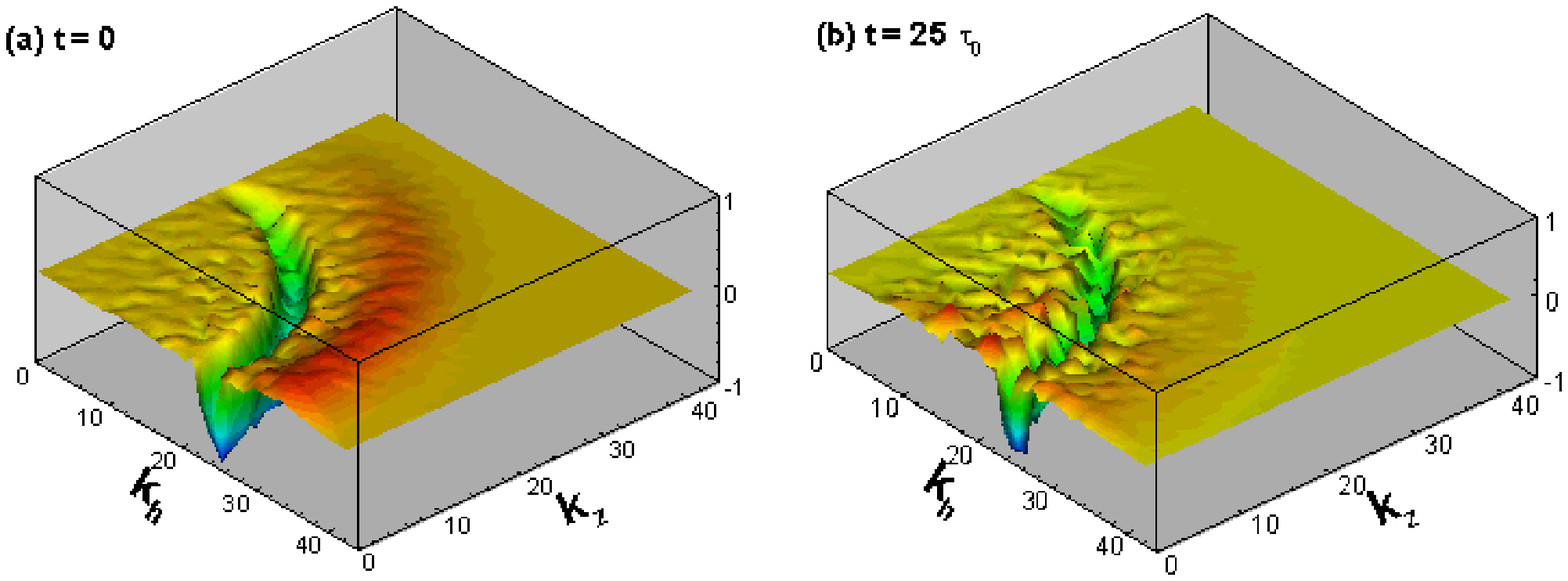,width=420pt,height=160pt}}
\centerline{\psfig{file=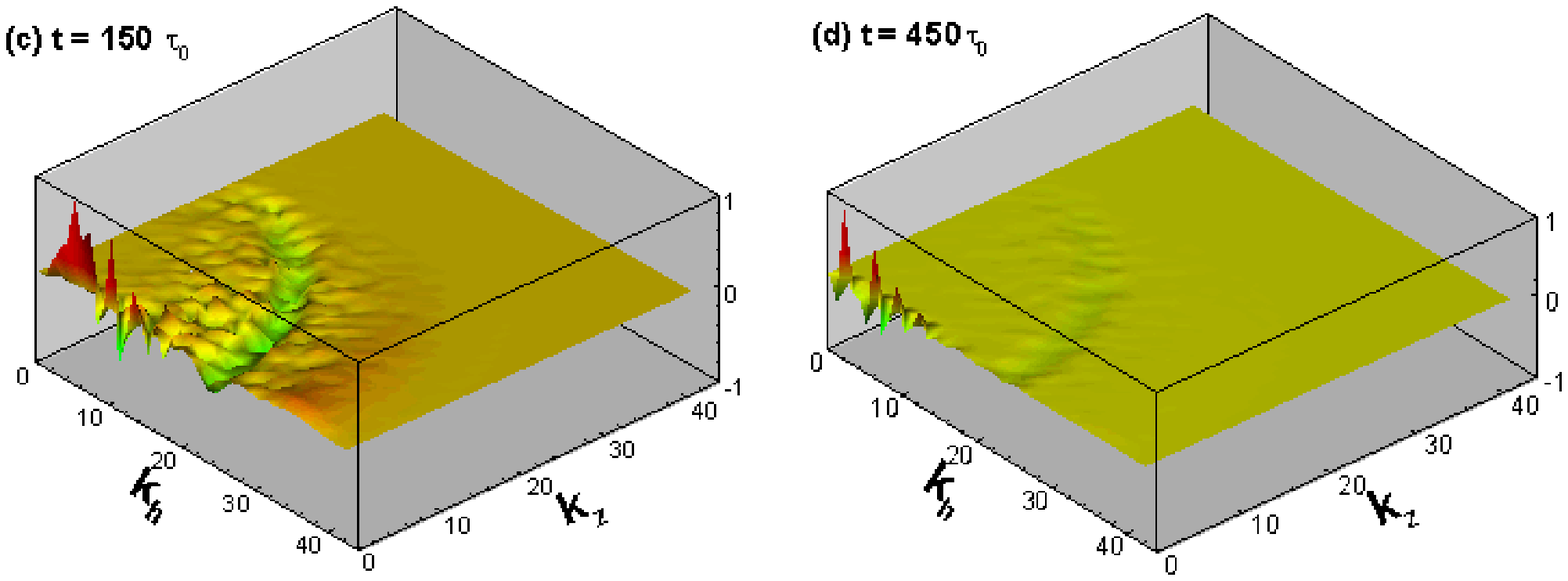,width=420pt,height=160pt}}
\noindent {\small FIGURE 2. Normalized energy transfer functions
$T(k_h,k_z)/|T|_{max}$ at different times for $Ro=0.066$. The red
color indicates the large positive transfer and the blue color
indicates the large negative transfer.}

\bigskip
\centerline{\psfig{file=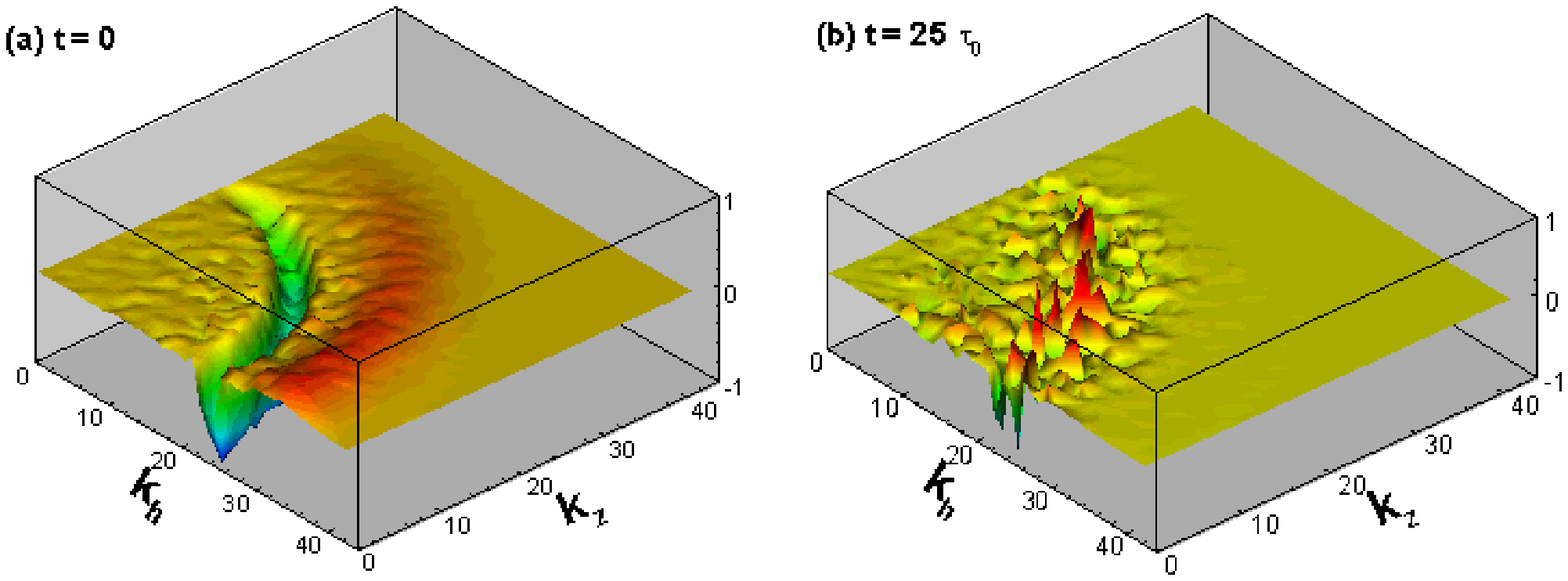,width=420pt,height=160pt}}
\centerline{\psfig{file=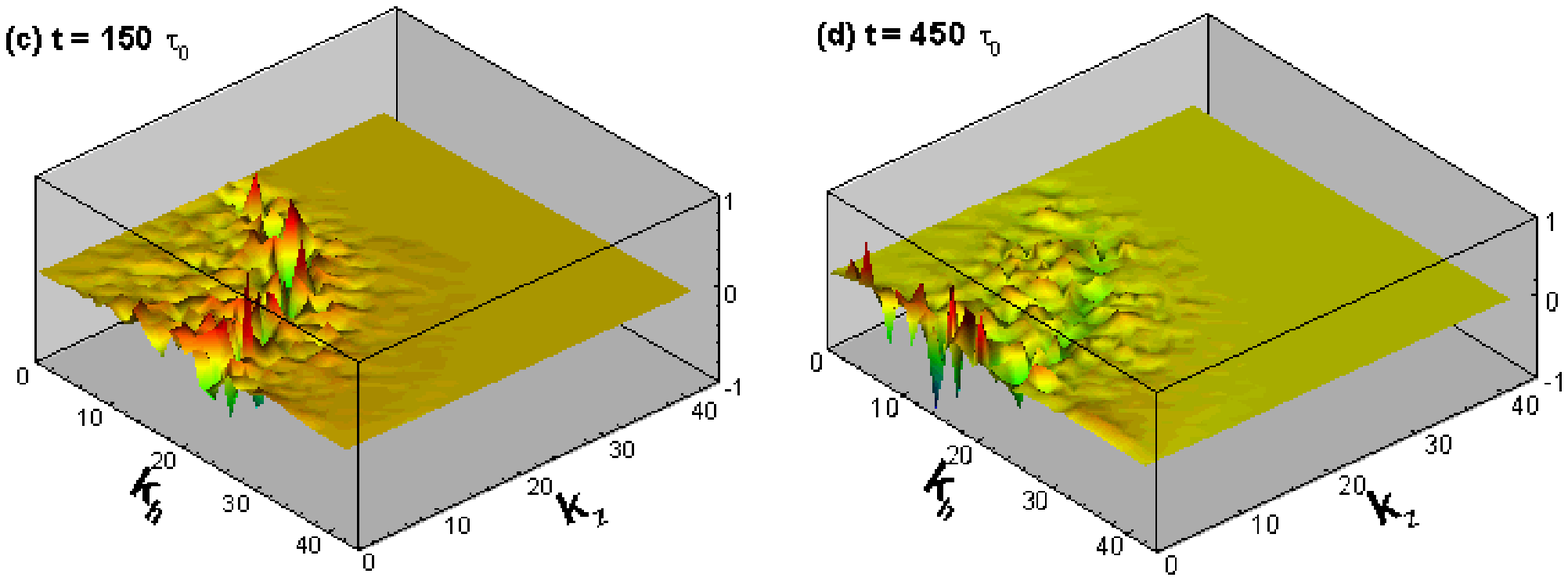,width=420pt,height=160pt}}
\noindent {\small FIGURE 3. Normalized energy transfer functions
$T(k_h,k_z)/|T|_{max}$ at different times for $Ro=0.0085$.}
\bigskip

\bigskip
\centerline{\psfig{file=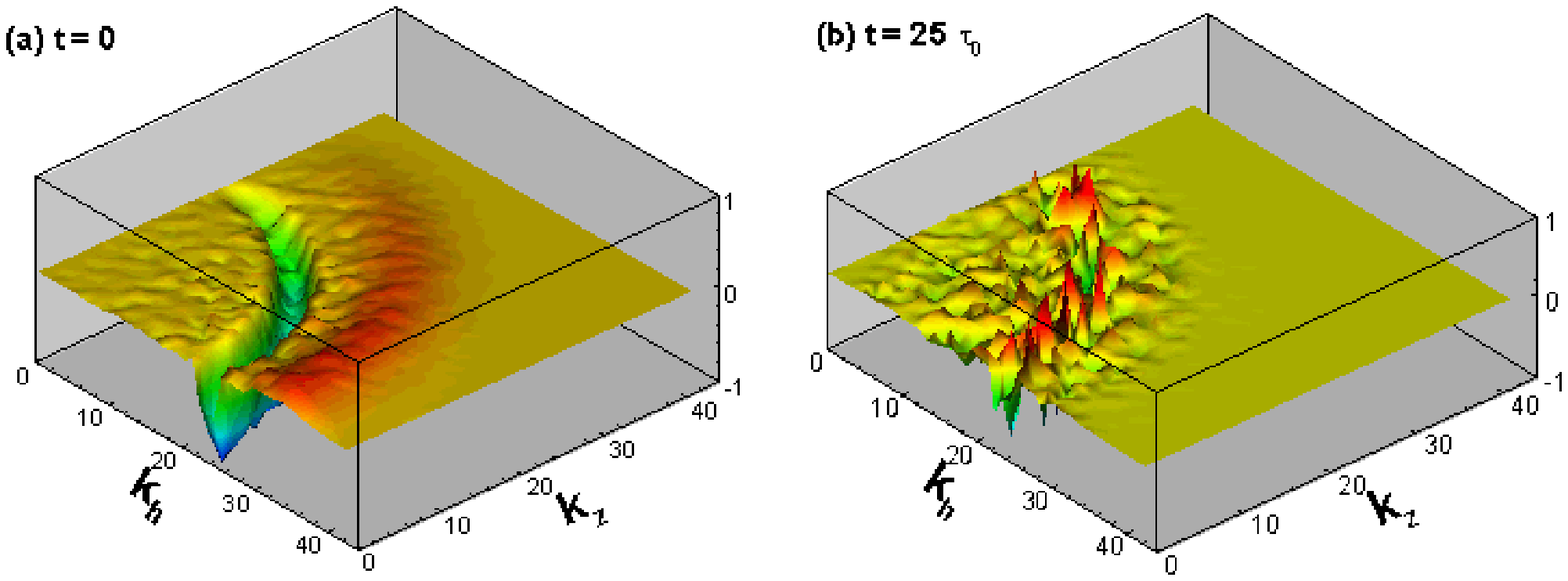,width=420pt,height=160pt}}
\centerline{\psfig{file=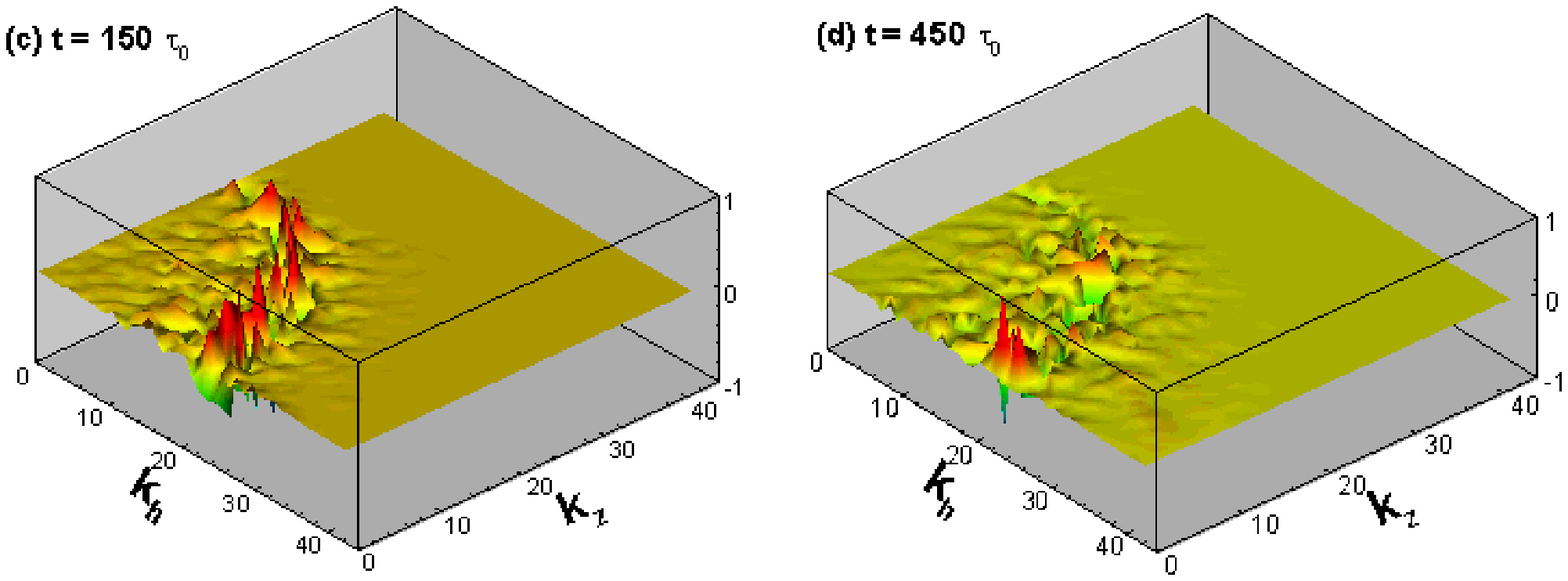,width=420pt,height=160pt}}
\noindent {\small FIGURE 4. Normalized energy transfer functions
$T(k_h,k_z)/|T|_{max}$ at different times for $Ro=0.0021$.}
\bigskip

The resonant wave theory predicts not just that
two-dimensionalization will occur, but also that the slow 2D modes
will satisfy 2D-3C dynamics, and, in particular, should exhibit
the phenomenology of the 2D inverse cascade. To test this, we
study the spectral transfer at various Rossby numbers. Energy is
conserved in detail for triads with all slow modes, so that we can
define their energy flux $\Pi_{sss}(k_h) = \int^\infty_{k_h}
T_{sss}(k_h')d k_h' $ where the energy transfer function
$T_{sss}(k_h)=\sum_{S(k_h)} Re[\hat{\bu}^* \bdot \hat{\bu} \times
\hat{\bom}]$ only accounts for the contribution from slow modes
$k_z=0$.\footnote{Before we calculate the transfer function
$T_{sss}(k_h)$, we need to zero out the velocity field for the
fast modes $k_z \neq 0$.} In Fig.~5 are plotted energy fluxes
$\Pi_{sss}(k_h)$ for different Rossby numbers together with the
energy flux $\Pi_{2D}(k_h)$ from the parallel 2D simulation. At
the larger Rossby number $Ro=0.066$, energy fluxes not only
fluctuate over time but also are much bigger than those at smaller
Rossby numbers. At $Ro=0.0021$, energy fluxes are negative at
large scales showing a complete inverse energy cascade range.
Moreover, they begin to develop a spectral range with constant
value, as expected in 2D turbulence, and in fact agree closely
with those from 2D-NS (Fig.~5 (c)). A consistent picture is
obtained by fitting power-laws to the energy spectra, at different
rotation rates. Fig.~6 (a) shows slow-mode energy spectra
$E(k_h,k_z=0)$ are closer to $k^{-3}$ than $k^{-5/3}$ at
$Ro=0.066,$ in agreement with the findings of \cite{LW99} at
comparable Rossby numbers. However, at $Ro=0.0021$ the spectrum
scales as $k^{-5/3}$ (Fig.~6 (b)) similar to that of the parallel
2D-NS simulation in Fig.~6 (c) and as expected for a transient 2D
inverse cascade with this forcing (\cite{SY93}).

\bigskip
\centerline{\psfig{file=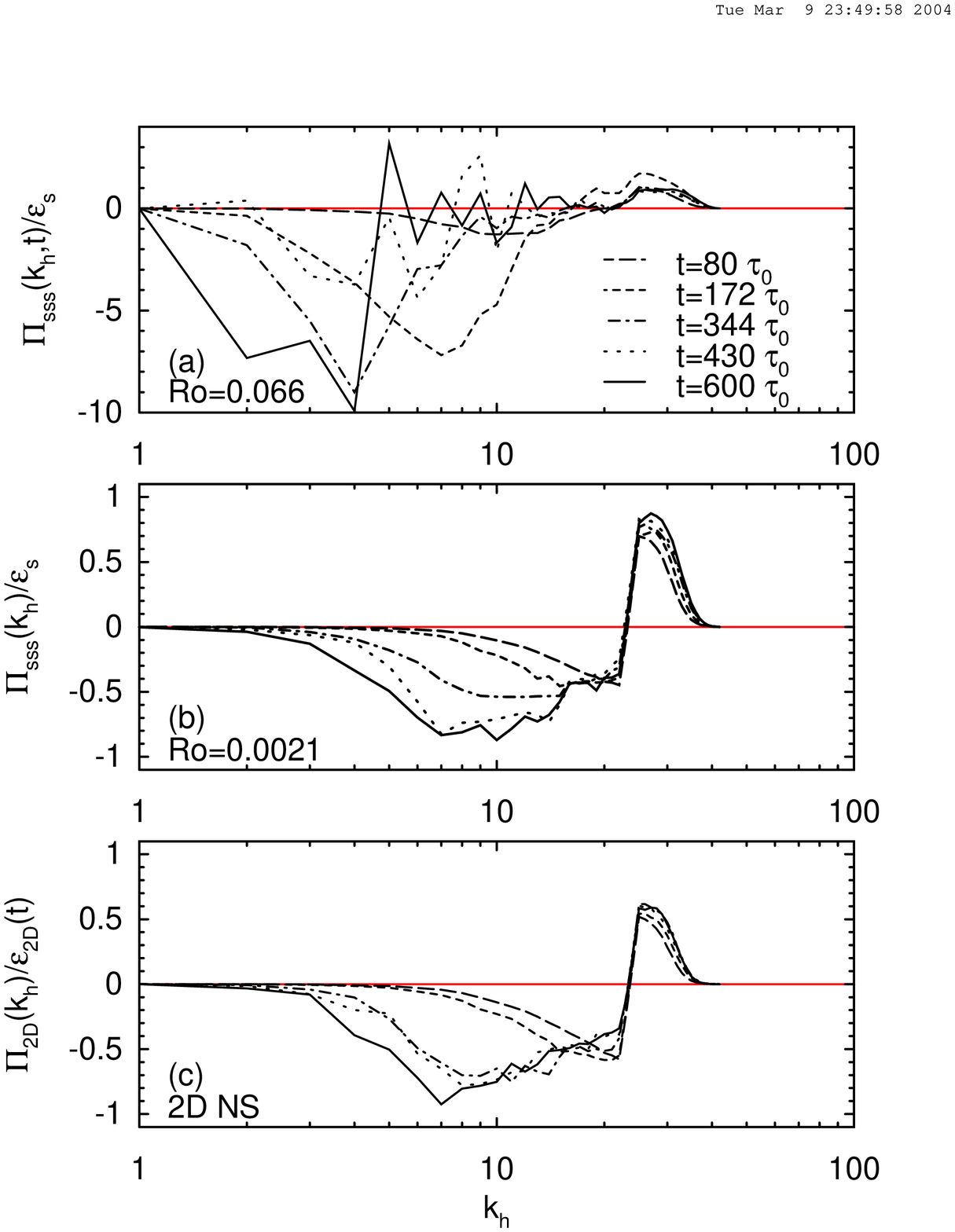,width=300pt,height=240pt}}
\noindent {\small FIGURE 5. Normalized energy fluxes from {\it
slow-slow-slow} interactions for (a)$Ro=0.066$; (b) $Ro=0.0021$;
(c) normalized energy flux from 2D Navier-Stokes at different
times. Here, $\epsilon_s$ is slow-mode energy input for 3D-NS with
rotation and $\epsilon_{2D}(t)$ is energy input for 2D-NS; $\tau_0
$ is the initial large eddy turnover time. $\epsilon_s$ is the
same for every rotation rate in our study. }
\bigskip

\bigskip
\centerline{\psfig{file=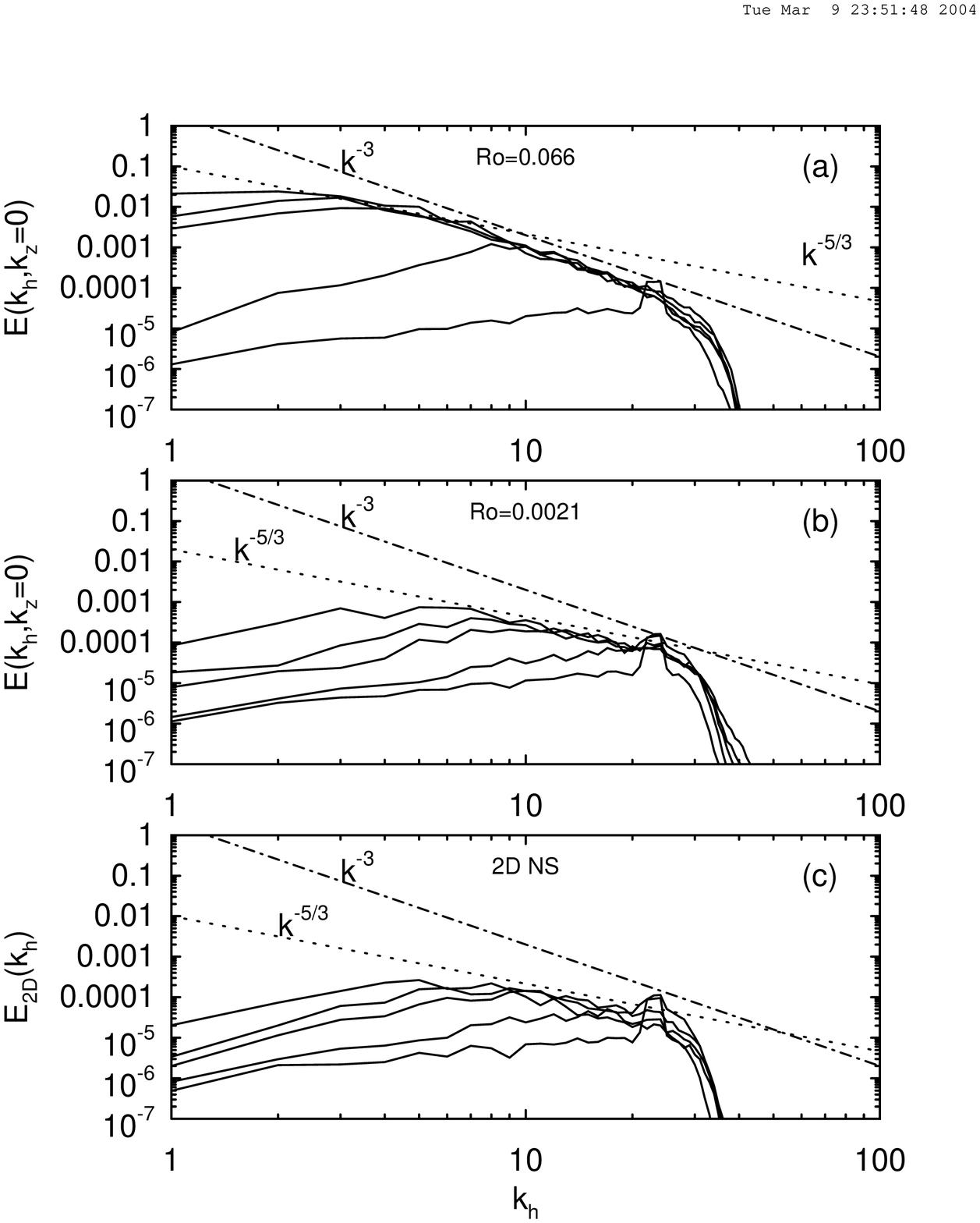,width=300pt,height=240pt}}
\noindent {\small FIGURE 6. Time evolution of energy spectra
(upward) from (a) slow modes at $Ro=0.066$; (b) slow modes at
$Ro=0.0021$; (c) 2D-NS at $t=80\tau_0$, $t=172\tau_0$,
$t=344\tau_0$, $t=430\tau_0$ and $t=600\tau_0$.}
\bigskip

\bigskip
\centerline{\psfig{file=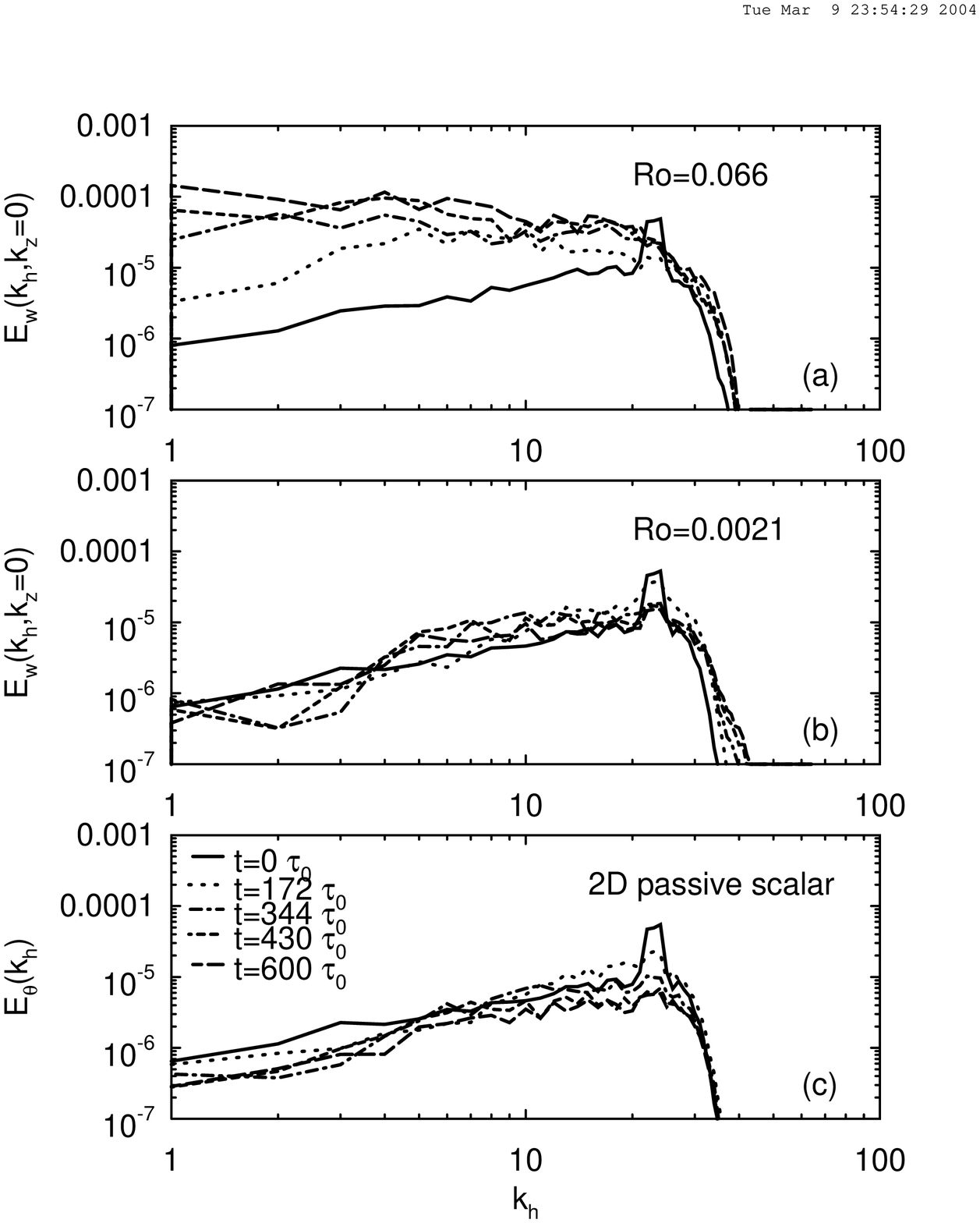,width=300pt,height=240pt}}
\noindent { \small FIGURE 7. Time evolution of energy spectra
(upward) of (a) $\overline{w}^{3D}$ at $Ro=0.066$; (b)
$\overline{w}^{3D}$ at $Ro=0.0021$ and (c) 2D passive scalar
$\theta^{2D}$.}
\bigskip

In Fig.~7 we present at various Rossby numbers the time evolution
of the energy spectrum $E_w(k_h, k_z=0)$ of the
vertically-averaged vertical velocity $\overline{w}^{3D},$
compared with the time evolution of the spectrum $E_\theta(k_h)$
of a passive scalar in the 2D parallel simulation. Fig.~7(a) shows
that at the highest simulated Rossby number $Ro=0.066$ there is a
tendency for an inverse cascade of $\overline{w}^{3D}$ towards the
large scales. This is opposite to what happens for a passive
scalar in a 2D inverse energy cascade range, which is well known
to experience a direct cascade to high wavenumber: see
\cite{CLMV00} and our own Fig.~7 (c). However, as $Ro$ decreases,
the energy spectra of $\overline{w}^{3D}$ become closer to those
of the 2D passive scalar, especially at the lowest simulated
Rossby number $Ro=0.0021$ (see Fig.~7 (b)). Thus,
$\overline{w}^{3D}$ tends also to transfer downscale as the
rotation rate increases.

The difference between the 3D slow-mode dynamics and 2D-NS at
finite Rossby numbers is also reflected in their different flow
structures. In Figs.~8--10 are plotted the iso-surfaces of the
vertically-averaged vertical vorticity $\overline{\omega}^{3D}_z$
from the 3D flow and those of the 2D vorticity $\omega^{2D}$.
Here, the averaged vertical vorticity is
$\overline{\omega}^{3D}_z=\frac{1}{H}\int^H_0 \omega_z(x,y,z)dz$.
In Fig.~8 the formation of large vortices is seen at the higher
Rossby number $Ro=0.066$. Both cyclonic and anticyclonic vortices
are present in the flow, which is different from the observations
from \cite{LW99}. In their simulations, only cyclonic vortices
were observed at larger Rossby numbers. At the same time, large
vortices are not visible in the 2D turbulent flow, which seems
still in its infancy (see Fig.~10). However, at a much smaller
Rossby number $Ro=0.0021$, the flow structures in 3D bear a strong
resemblance to those seen in 2D (see Fig.~9 and Fig.~10).

The Dynamic Taylor-Proudman Theorem predicts that the 3D slow
modes will obey the 2D-3C Navier-Stokes equations, asymptotically
for low Rossby numbers. As we discussed in the previous section,
\cite{EM96} and \cite{ME98} measured the size of the deviations
from this prediction by Sobolev norms of the error field with
large $p$, which do not yield useful estimates for high Reynolds
number flow. Therefore, we shall consider here instead $p=0$, or
the error energy itself, as in \cite{Kr70} and \cite{Leith72}. It
is useful to divide this error energy into separate contributions
from horizontal and vertical velocity components. Precisely, we
define
\be E_{\delta,H}(t)=\frac{1}{2} \int d^2\bx\,\,|\overline{{\bf
u}}^{3D}_H-\bu^{2D}|^2,\,\,\,\, E_{\delta,V}(t)=\frac{1}{2} \int
d^2\bx \,\,|\overline{w}^{3D}-\theta^{2D}|^2. \lb{error-en} \ee
As before,
$\overline{\bu}^{3D}_H=(\overline{u}^{3D},\overline{v}^{3D})$ and
$\overline{w}^{3D}$ are the vertically-averaged horizontal and
vertical velocities, respectively. For comparison,
$\bu^{2D}=(u^{2D},v^{2D})$ is the solution of the 2D-NS equation
(\ref{2D_N-S}) and $\theta^{2D}$ is the solution of the 2D passive
scalar equation (\ref{2D_passive}). In Fig.~11 (a) and (b) are
shown the error energies in (\ref{error-en}), normalized by the
energies $E_H(t,Ro),E_V(t,Ro)$ of the corresponding 3D
vertically-averaged fields, as functions of time at different
Rossby numbers. Both plots show that the normalized error energy,
at least over a finite interval of time, decreases as Rossby
number is lowered. For any finite value of $Ro,$ the
vertically-averaged 3D solutions and the 2D solutions begin to
diverge as time $t$ increases because of their chaotic dynamics
and become uncorrelated, leading to saturation of the normalized
error energy at sufficiently long times. Although presently
available theorems do not rigorously imply the behavior observed
in Fig.~10 for the $p=0$ norms, nevertheless these results are
quite in line with expectations from the formal multi-time scale
analysis (\cite{Green68,Waleffe93}).

An important quantity to determine is the maximal length $T_*$ of
the slow-time interval over which the resonant wave theory becomes
valid as $Ro\rightarrow 0$ . The rigorous theorems guarantee that
a finite time $T>0$ exists so that Sobolev norms of the error
(with $p>1+d/2$) converge to zero for all slow times $t\in [0,T]$.
However, the asymptotic analysis does not determine the largest
possible time $T$ for which this is true. In order to estimate
this, we have defined $T(Ro)$ for each Rossby number $Ro$ as the
time $t$ for which the normalized error energy plotted in
Fig.~10(a) first reaches the value $0.5$. This quantity is plotted
as a function of $Ro$ in Fig.~12. For a wide range of Rossby
numbers above 0.00825, the graph can be fit by a power-law
$Ro^{-1/2}$ with a small error bar (only $2\%$). The reason for
such a power law to exist in this range is not clear.
Interestingly, there appears to exist a transition when Rossby
number is below 0.00825 and the graph tends to saturate to a
constant. If true, this implies that the resonant wave theory can
be valid, even for arbitrarily small $Ro,$ only in a finite time
interval of length $<T_*=\lim_{Ro\rightarrow 0}T(Ro).$

\smallskip

\centerline{\psfig{file=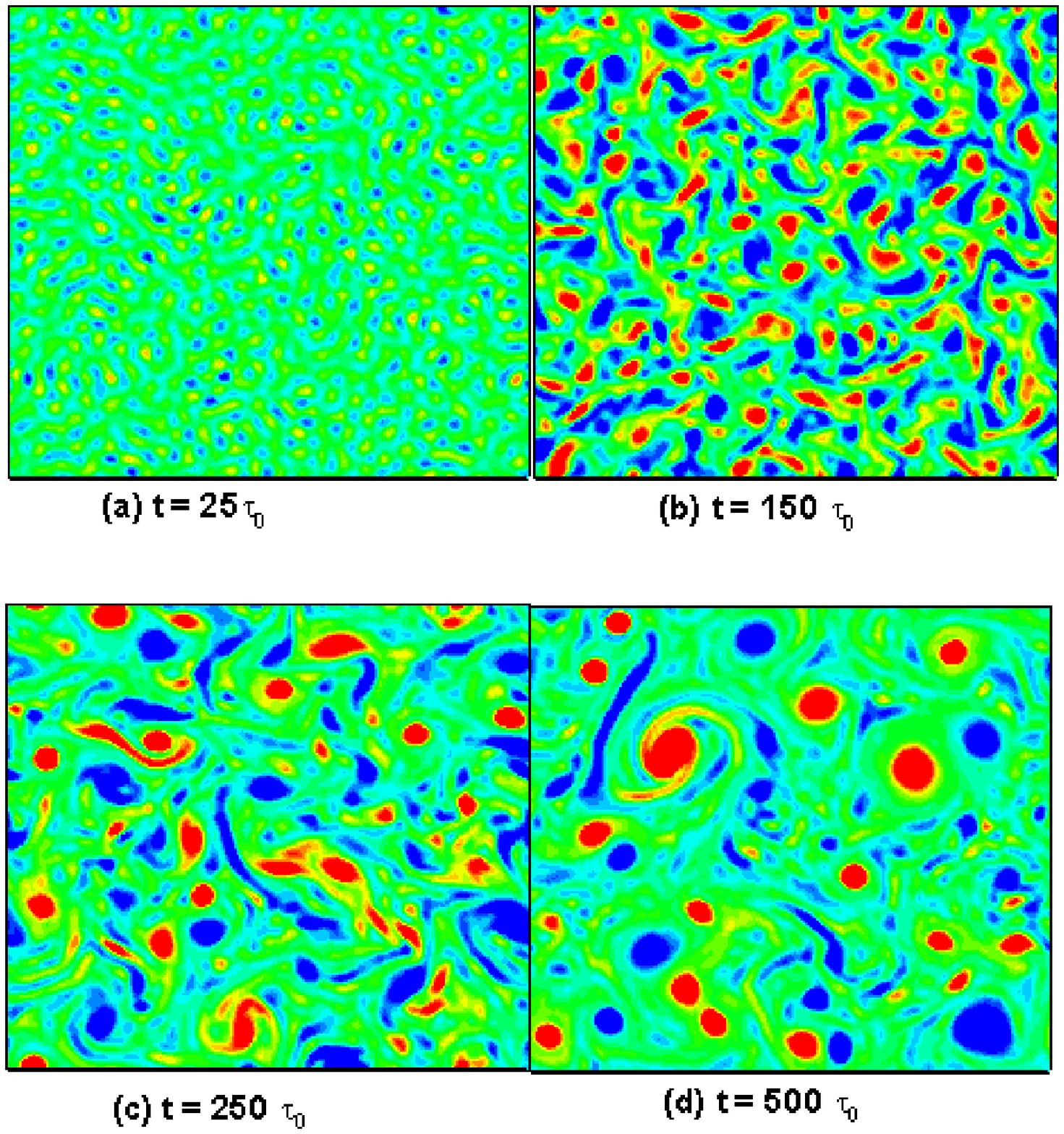,width=300pt,height=250pt}}
\noindent {\small FIGURE 8. Time histories of the
vertically-averaged vertical vorticity $\overline{\omega}^{3D}_z$
iso-surfaces from 3D rotating turbulence at $Ro=0.066$. The red
color is for the large positive vorticity and the blue one is for
the large negative vorticity.}

\smallskip
\centerline{\psfig{file=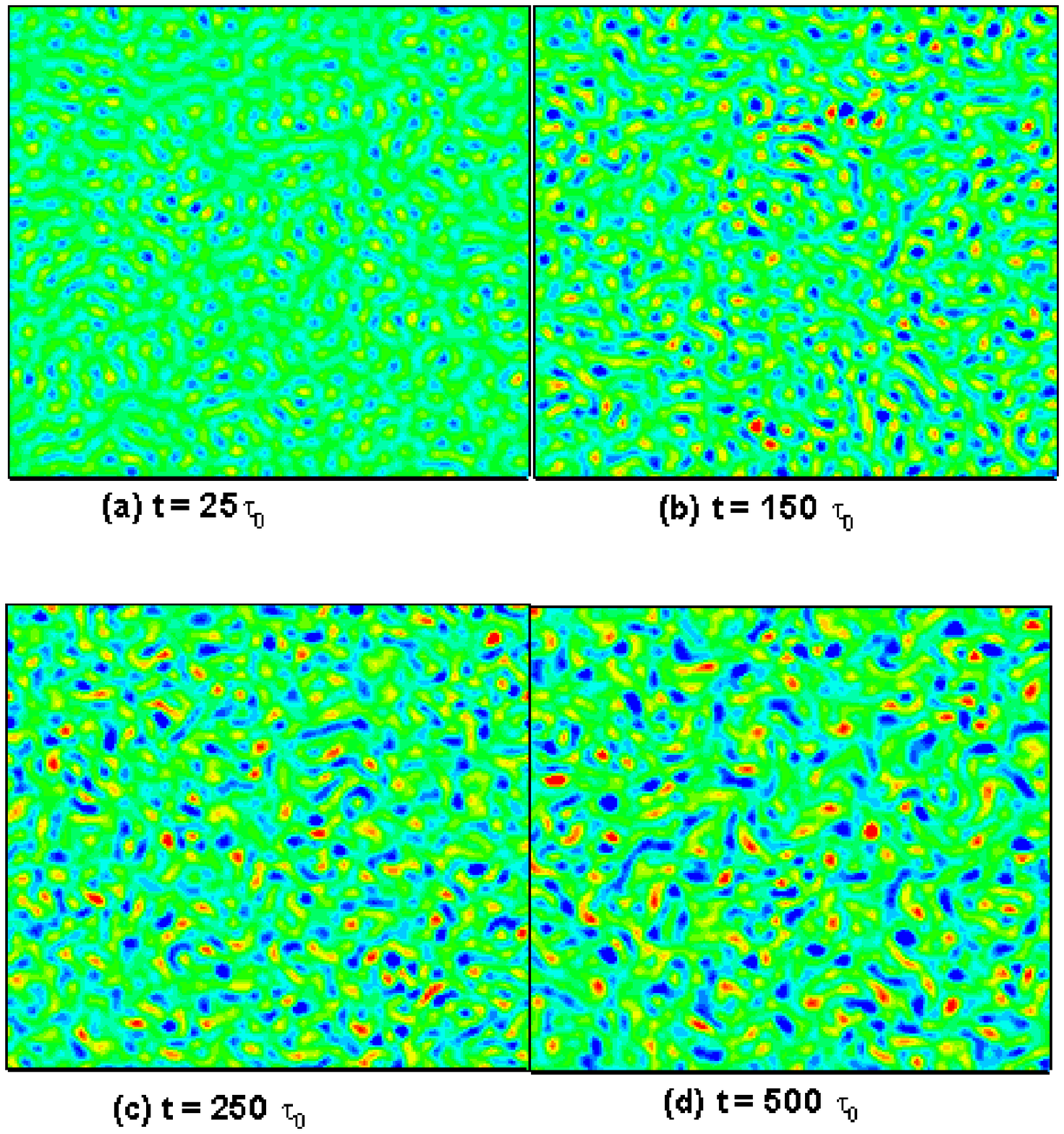,width=300pt,height=250pt}}
\noindent {\small FIGURE 9. Time histories of the
vertically-averaged vertical vorticity $\overline{\omega}^{3D}_z$
iso-surfaces from 3D rotating turbulence at $Ro=0.0021$.}
\smallskip

\smallskip
\centerline{\psfig{file=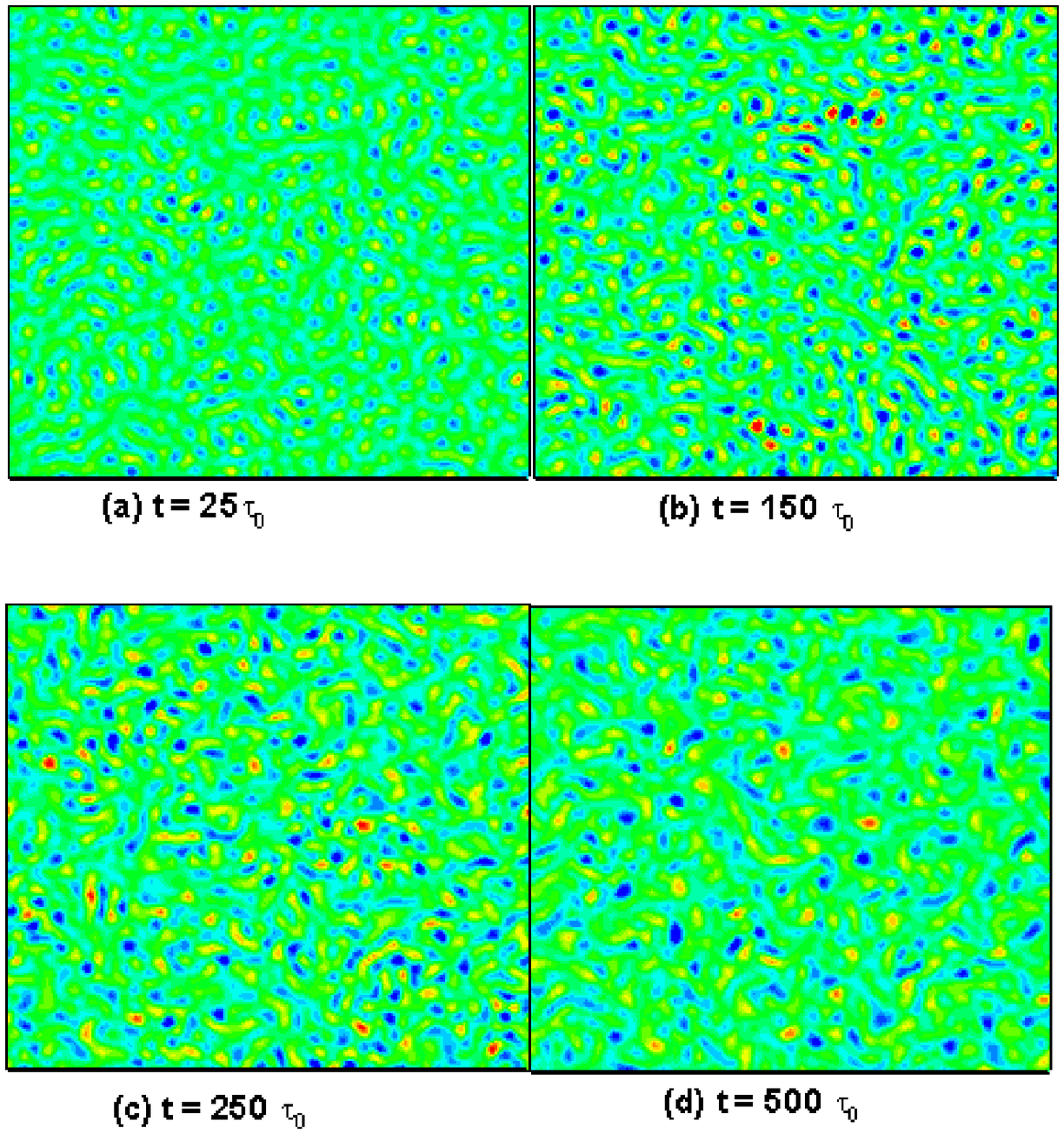,width=300pt,height=250pt}}
\noindent {\small FIGURE 10. Time histories of the $\omega_z^{2D}$
vorticity iso-surfaces from 2D-NS.}
\smallskip

\bigskip
\centerline{\psfig{file=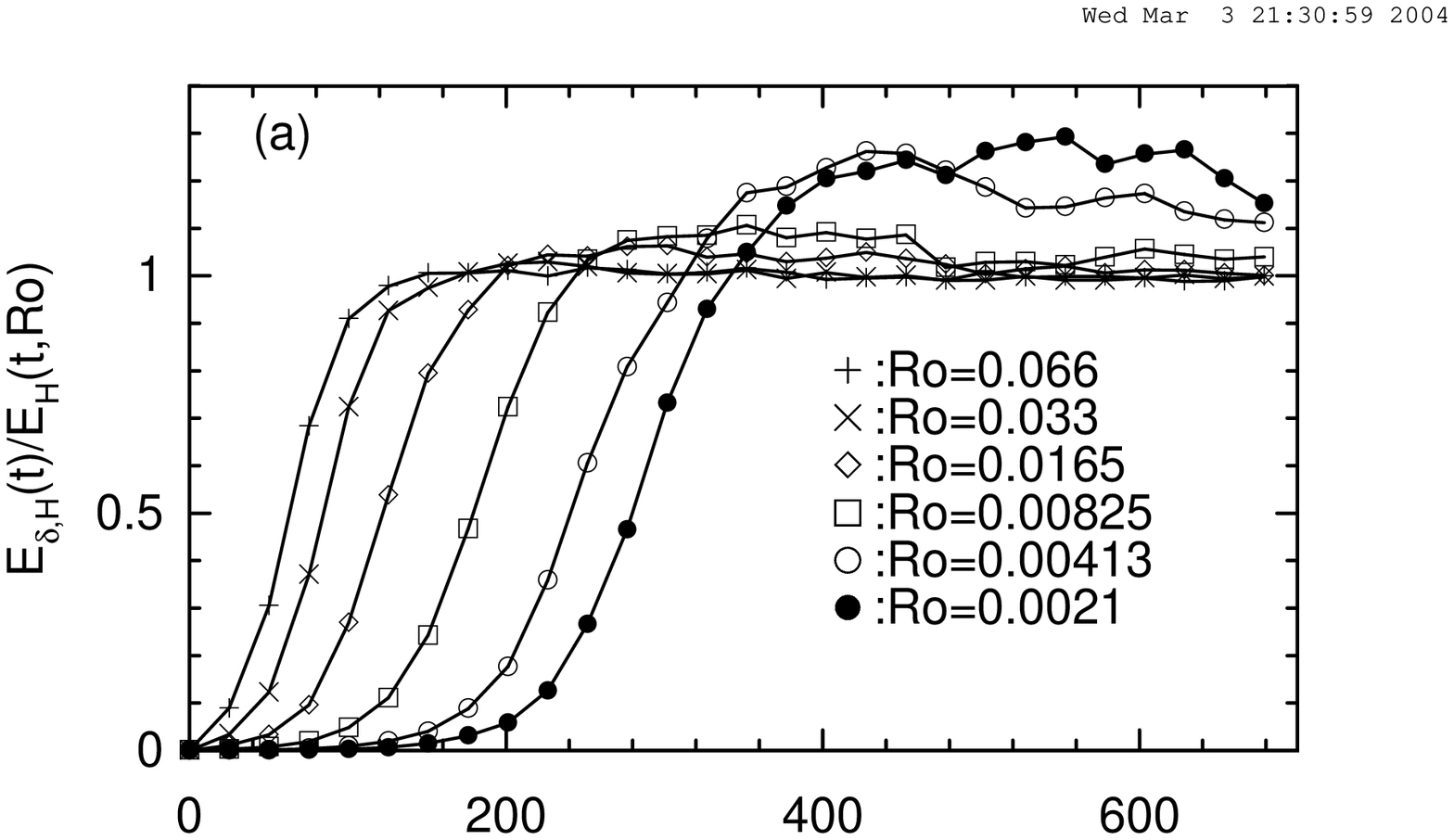,width=300pt,height=150pt}}
\centerline{\psfig{file=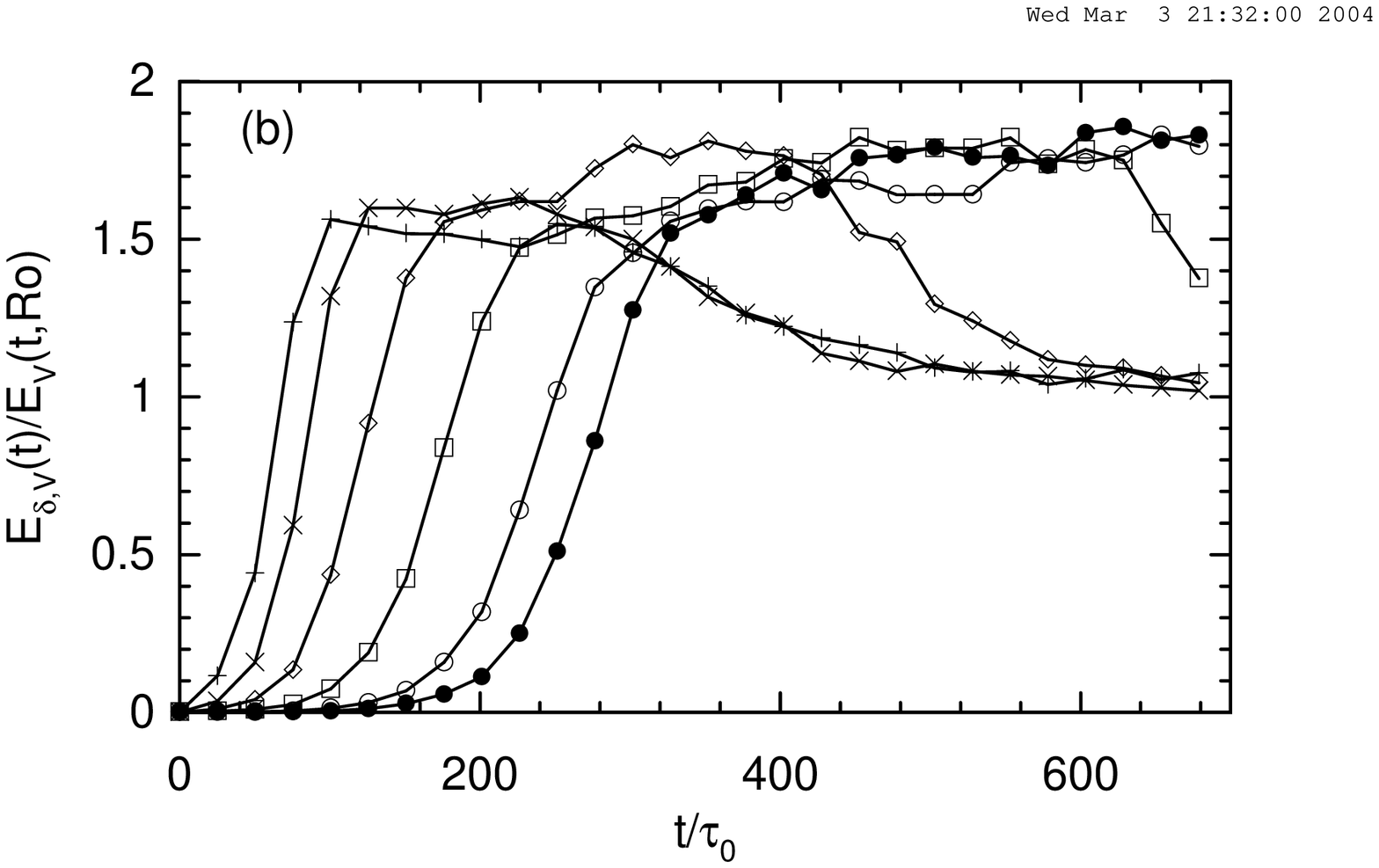,width=300pt,height=150pt}}
\noindent {\small FIGURE 11. Normalized error energy for (a)
$\overline{\bu}^{3D}_H-\bu^{2D}$ and (b)
$\overline{w}^{3D}-\theta^{2D},$ as functions of time at different
Rossby numbers. Here $E_H(t,Ro)=\frac{1}{2}\int
|\overline{\bu}^{3D}_H|^2$ and $E_V(t,Ro)=\frac{1}{2}\int |
\overline{w}^{3D}|^2$.}
\bigskip

\bigskip
\centerline{\psfig{file=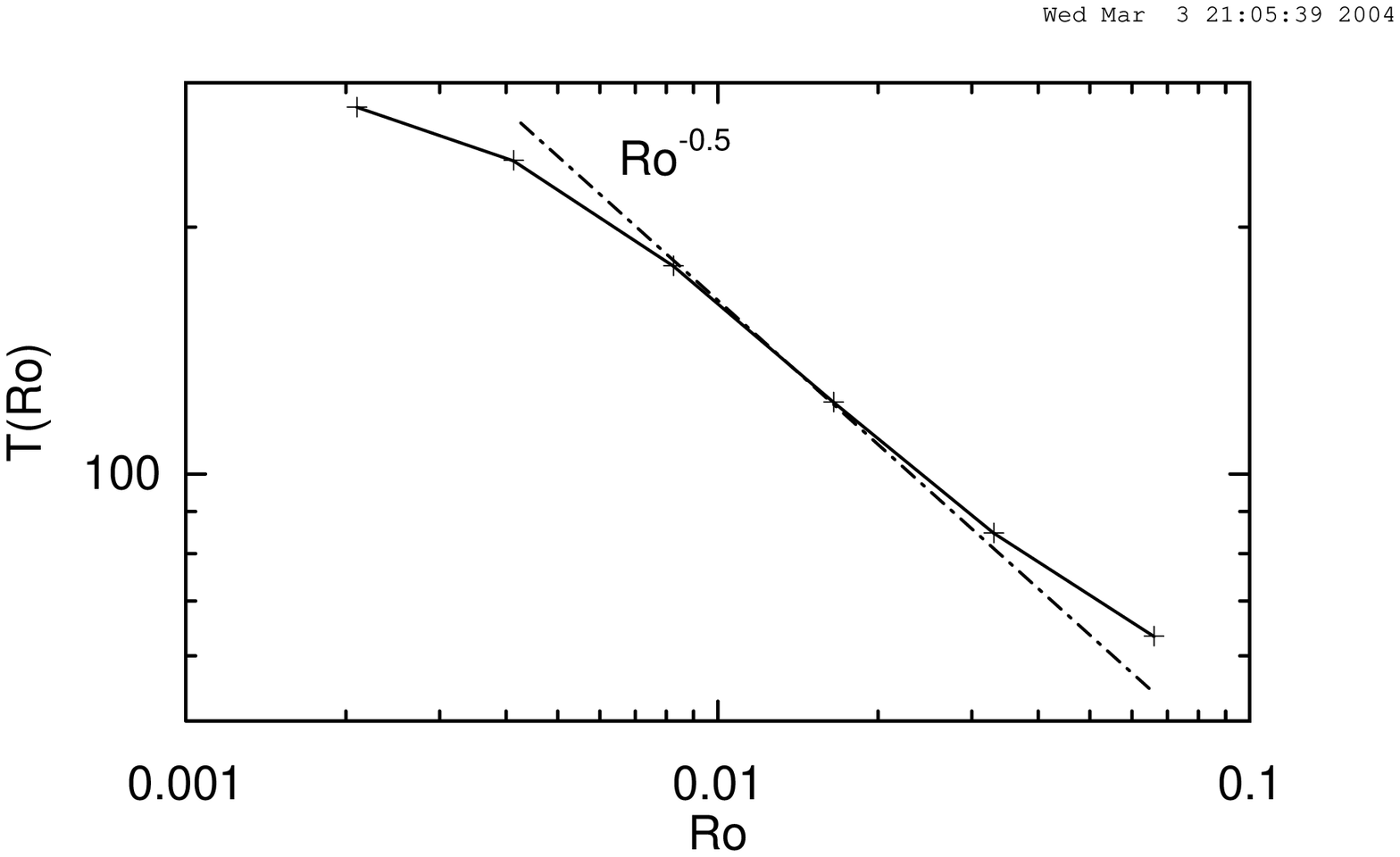,width=300pt,height=150pt}}
\noindent {\small FIGURE 12. Time $T(Ro)$ vs. Rossby number $Ro$.
There exists a power law scaling $Ro^{-0.5}$ in the range
$Ro>0.00825$ but the graph appears possibly to platueau below
$Ro=0.00825$. }
\bigskip

These findings are consistent with the expected segregation of the
autonomous 2D slow modes from the 3D fast modes at low Rossby
numbers. All ``slow-fast-fast'' triads transferring energy from a
slow mode by interactions with two fast modes are non-resonant and
it is another main prediction of the resonant wave theory that
such transfer should disappear as $Ro \rightarrow 0.$ To test
this, we calculate an energy flux into low-wavenumber slow modes
from slow-fast-fast triadic interactions. Precisely, we define
$\Pi_{sff}(k_h)= -\int^{k_h}_0 T_{sff}(k',k_z=0)dk'$ where
$T_{sff}(k_h,k_z=0)=T(k_h,k_z=0)-T_{sss}(k_h)$ is the energy
transfer function from fast modes into slow modes at horizontal
wavenumber $k_h$. Fig.~13 (a-c) show that energy flux into slow
modes at small $k_h$ decreases as $Ro$ decreases. In another
words, less and less fast-mode energy is transferred into the
large scales in the 2D plane as $Ro\rightarrow 0$. At larger
Rossby numbers, more energy is drained from fast modes directly
into the large-scale slow modes by such non-resonant transfer,
causing the flow to become two-dimensional more quickly than at
smaller Rossby numbers. \cite{LW99} suggested that this
non-resonant mechanism was responsible for the rapid
two-dimensionalization in their simulation at larger Rossby
numbers. \footnote{Although we agree with their conclusion, we do
not find the argument they offered very convincing. In their
simulation, the forcing was white-noise in time with a wavenumber
spectrum $F(k)=\epsilon_k exp(-0.5(k-k_f)^2/\sigma^2)/\sqrt{2\pi
\sigma^2},$ nonzero in a spherical shell with mean radius $k_f$
and width $\sigma.$ They attributed the strong
two-dimensionalization they observed to non-resonant interactions,
because of what they claimed was purely three-dimensional forcing.
However, their forcing input energy not only into the fully
three-dimensional modes but also into two-dimensional modes in the
circular band where the spherical shell intersects the $k_z=0$
plane. This energy will inverse cascade to large scales by
``slow-slow-slow'' resonant interactions.}

\bigskip
\centerline{\psfig{file=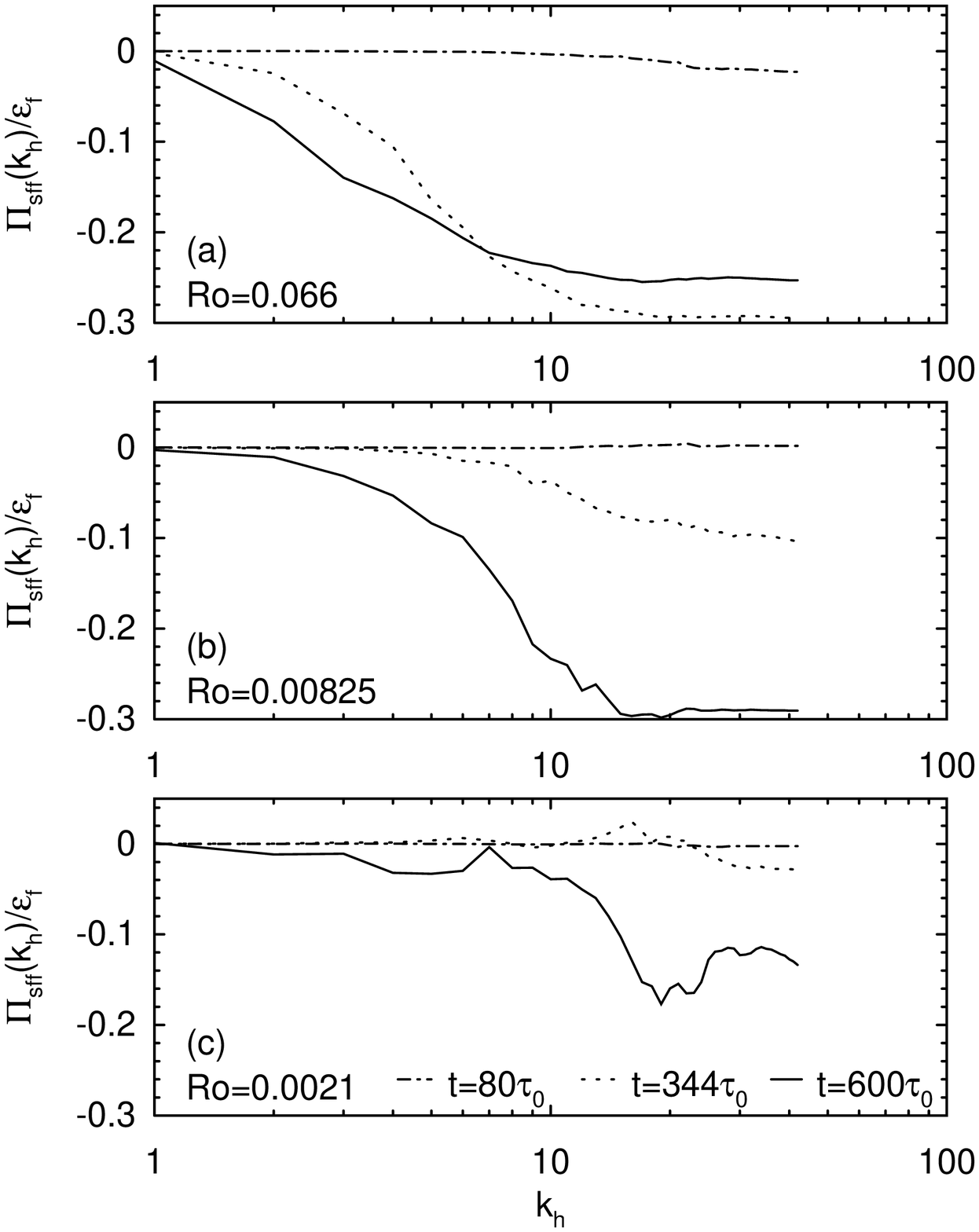,width=300pt,height=240pt}}
\noindent {\small FIGURE 13. Normalized energy fluxes
$\Pi_{sff}(k_h)$ from {\it slow-fast-fast} interactions for (a)
$Ro=0.066$; (b) $Ro=0.00825$; (c) $Ro=0.0021$ at different times.
Here, $\epsilon_f$ is the total energy input into the fast 3D
modes.}
\bigskip

A final prediction of the resonant wave theory is that, as $Ro
\rightarrow 0$, fast-mode energy will tend to be transferred
toward smaller values of $\cos \theta$ by ``fast-fast-fast''
resonant interactions. \cite{Waleffe93} has demonstrated this
tendency using the resonance condition (\ref{resonance}) and a
plausible ``instability hypothesis,'' which asserts that energy
will tend to be transferred out from the unstable leg of a
wavevector triad. \cite{Mansour92} and \cite{Cambon89} have
observed that energy was transferred towards small $\cos \theta$.
However, neither distinguished fast-modes from slow-modes. Thus,
the effects they saw may also be due to non-resonant transfer from
fast modes directly into slow-modes. In order to investigate the
``fast-fast-fast'' interactions, we calculate the energy
distribution $E_f(\cos\theta,t)$ only from fast modes. (We recall
that the catalytic ``fast-slow-fast'' interactions can only
transfer energy between fast modes with the same value of
$\cos\theta$ and thus cannot contribute to the dynamics of this
quantity.) Fig.~14 (a) shows that $E_f(\cos\theta, t)$ flattens
monotonically with time at the largest Rossby number $Ro=0.066$,
implying fast modes behave largely as three-dimensional. At the
middle Rossby number $Ro=0.00825$ (Fig.~14 (b)), $E_f(\cos\theta,
t)$ peaks at smaller $\cos\theta$ at earlier times and then
flattens everywhere at later times. However, at the smallest
Rossby number $Ro=0.0021$, fast-mode energy monotonically builds
up at small $\cos \theta$ (Fig.~14 (c)). This supports the
conclusions of \cite{Waleffe93} based on the ``instability
hypothesis''.

\bigskip
\centerline{\psfig{file=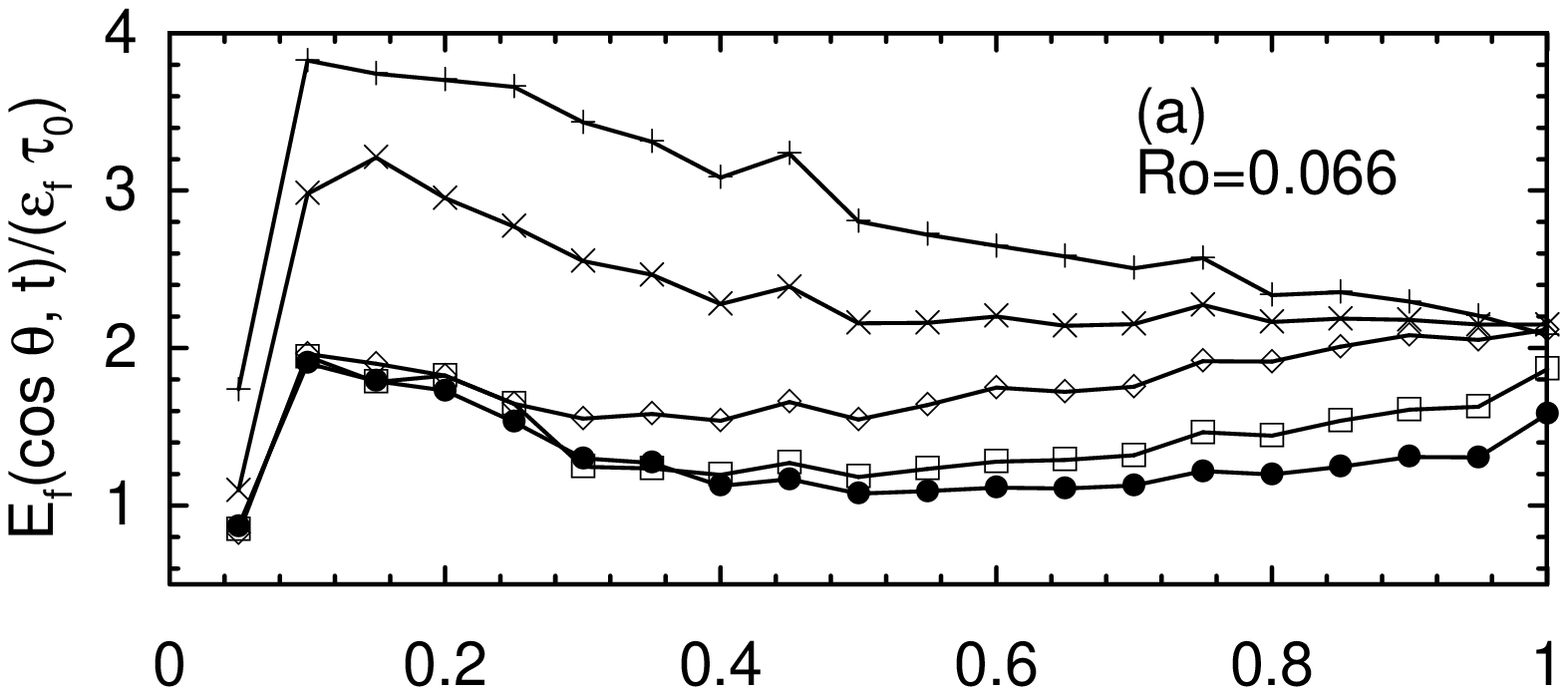,width=300pt,height=80pt}}
\centerline{\psfig{file=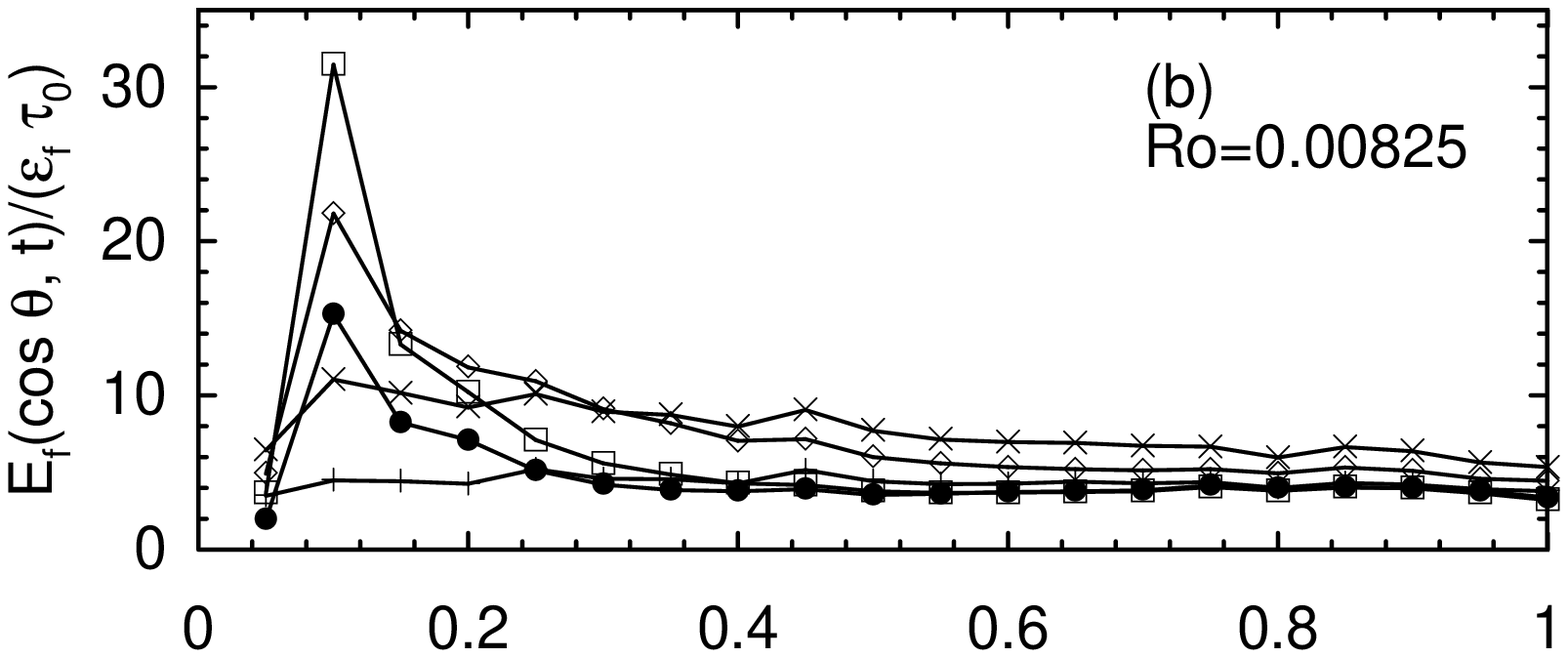,width=300pt,height=80pt}}
\centerline{\psfig{file=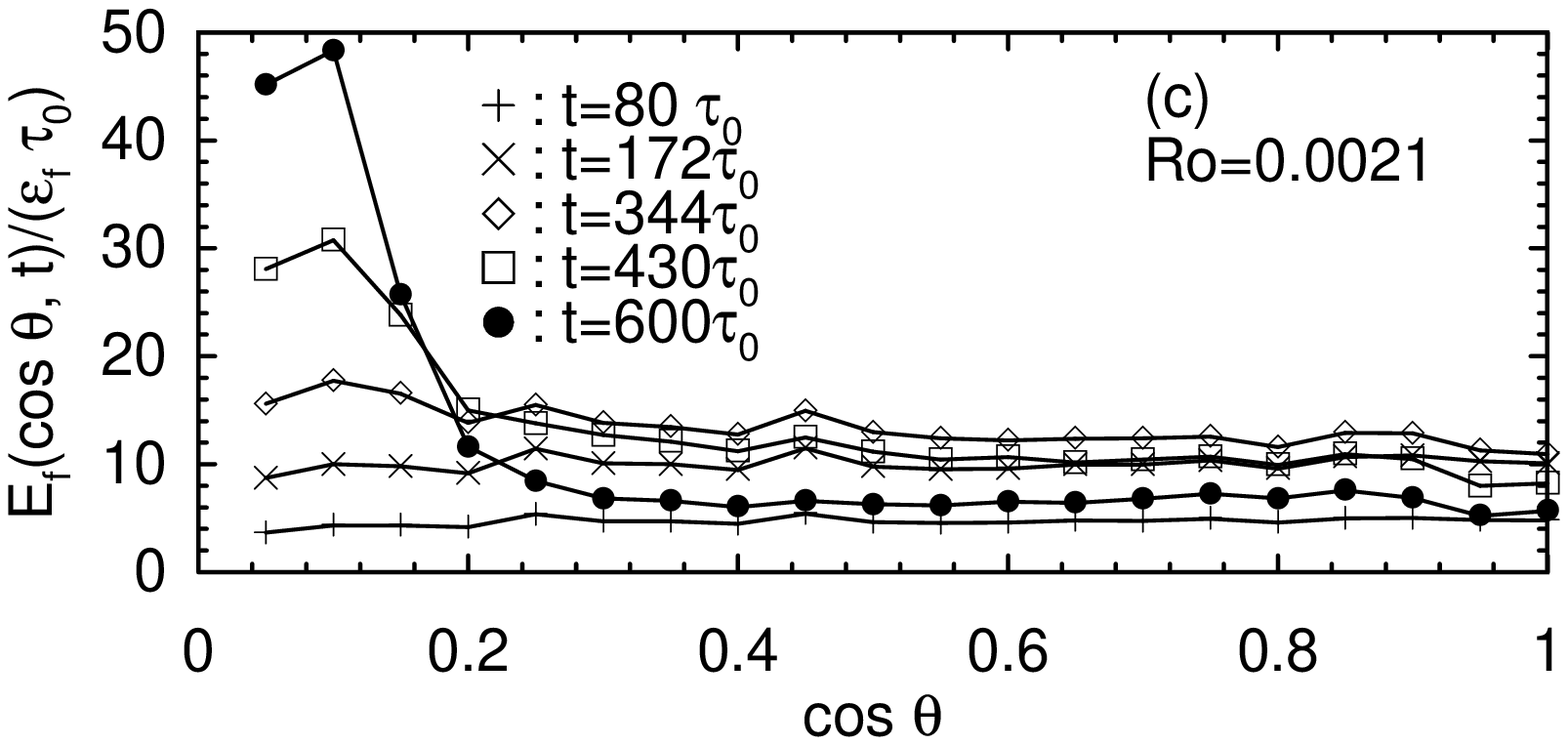,width=300pt,height=80pt}}
\noindent {\small FIGURE 14. Normalized fast-mode energy
distributions
 $E_f(cos\theta,t)/(\epsilon_f \tau_0)$ for (a)  $Ro=0.066$;(b)
$Ro=0.00825$; (c) $Ro=0.0021$ at different times. }
\bigskip

\section{Conclusions}

To summarize briefly our results, this is the first numerical
simulation of forced homogeneous rotating turbulence to explore
quantitatively the long-time effects of resonant interactions on
the two-dimensionalization as $Ro \rightarrow 0 $. We demonstrate
an inverse cascade regime with the characteristics of 2D
turbulence, namely, a $k_h^{-5/3}$ spectral range where energy
flux is negative and constant for the slow-mode dynamics. This is
consistent with the predictions of 2D-3C equations, corresponding
to resonant interactions. Three additional findings verify the
increasing importance of resonant interactions to
two-dimensionalization at small $Ro$. First, vertically-averaged
3D velocities $(\ou^{3D},\ov^{3D},\ow^{3D})$ and the solutions of
2D-3C equations converge as $Ro \rightarrow 0,$ in the sense that
the energy in their difference fields vanishes. Second,
non-resonant energy flux into small $k_h$ in the $k_z=0$ plane
from slow-fast-fast triads decreases as $Ro$ decreases. Finally,
fast-mode energy is transferred toward the $k_z=0$ plane,
consistent with the prediction of \cite{Waleffe93} that resonant
interactions of fast modes drive the flow to become quasi-2D.

Our work has investigated rapidly-rotating turbulent fluids as a
case study in the application of resonant wave theory.  The
results of the study therefore have more general interest than for
this particular problem. The same type of theory has been widely
applied to geophysical turbulent flows with rotation and/or
stratification (\cite{Phillips81,Craik85}). Details of the
predictions may differ considerably according to the context. For
example, \cite{Bartello95} used a similar reasoning as that of
\cite{Waleffe93} to argue that for rotating, stratified turbulence
the catalytic, ``fast-slow-fast'' resonant triads will transfer
the energy in fast gravity waves downscale and provide a primary
mechanism of nonlinear geostrophic adjustment, while the
``fast-fast-fast'' resonant interactions will be of secondary
importance. In rapidly rotating fluids without density
stratification, we find that the opposite is true. Nevertheless,
many of the basic issues are the same in all applications of the
resonant wave theory to high Reynolds-number turbulent flows.
Therefore, it may be useful to summarize several key conclusions
from our investigation of this one example.

First, the main predictions of the wave-resonance theory appear to
hold for $Ro\ll 1$ and $Re\gg 1$ simultaneously. Convergence of
the approximation holds in the useful sense that the energy in the
error fields decreases to zero, over a finite time-interval.
Present rigorous theorems (\cite{EM96,EM98,ME98}) use Sobolev-norm
estimates that yield useful estimates for rotating laminar flows,
but not for fully-developed turbulence. An optimistic conclusion
of our work is that those theorems can probably be improved, by
substitution of an $L^2$ or energy norm for the Sobolev norms. A
proof of this may not be easy, given the current lack of
understanding of the inviscid, high-Reynolds number limit of fluid
equations. However, our numerical results lend support to the idea
that the energy-containing, large-scale motions in turbulent flow
will conform to the predictions of resonant wave theory in the
appropriate limit.

We find that there is an intrinsic time restriction on the
validity of the resonant wave theory. The formal multiple-scale
analysis, as well as the rigorous theorems, predict that there
should be a finite interval of the slow-time variable $t$ over
which the approximation will converge. They do not answer the
question whether this time-interval is finite in fact, or whether
the approximation will be valid at any arbitrary time for a
sufficiently small Rossby number. Our numerical results seem to
indicate either that the convergence holds over only a finite
time-interval or else that the Rossby numbers required for a close
approximation become extremely small beyond a certain
time-horizon. Of course, both the  dynamics involved---rotating
3D-NS (\ref{1_Dimen-N-S}) and the low Rossby-number limiting
dynamics given by the averaged equation (\ref{averaged})---are
chaotic at high Reynolds number and a rapid exponential divergence
of their solutions is to be expected for finite $Ro$. This does
not imply a necessary failure of the resonant wave theory applied
to forced, statistically stationary turbulence, since chaotic
dynamics often possess ``structural stability'' of their long-time
statistics.

A more serious limitation of the wave-resonance theory may be the
magnitude of the Rossby numbers required for convergence, even at
relatively early times. We have found sizable non-resonant effects
for Rossby numbers as small as $Ro=0.066.$ Since the Rossby number
is not extremely small in most engineering applications, flow
two-dimensionalization is there probably due mainly to the
non-resonant interactions, for which the 2D dynamics is not
segregated from the 3D dynamics. However, the resonant wave theory
provides the correct ``rapid-distortion limit'' for fast
rotations. Therefore, it may still be useful to take into account
the resonance conditions in statistical modeling schemes for
engineering purposes, as a constraint which becomes exact as
$Ro\rightarrow 0.$ Resonant interactions may be more important for
geophysical flows, where typical Rossby numbers are much
smaller. However, for the atmosphere at midlatitude
synoptic-scales, the typical Rossby number is only about $Ro=0.1$
and even for the Gulf Stream with smaller characteristic
velocities the Rossby number is only $Ro=0.07$
(\cite{Pedlosky87}). In our simulations, this is a marginal value
to observe the predictions of the theory. A combination of
rotation with other effects such as stratification may be
necessary in the atmosphere and oceans to select resonant
interactions of fast waves.

\vspace{.2in}

{\bf Acknowledgements.} Direct numerical simulations are performed
in LSSC-II at the State Key Laboratory on Scientific and
Engineering Computing in China and at the cluster computer
supported by NSF Grant No. CTS-0079674 at the Johns Hopkins
University.

\appendix
\section{Dynamic Taylor-Proudman Theorem}
Here, we use the helical decomposition to give a simpler and
self-contained derivation of the Dynamic Taylor-Proudman Theorem
for the ``slow-slow-slow'' resonant triadic interactions. In a
triad consisting of all slow modes $\bk$,$\bp$ and $\bq$, then
$k_z=0$, $p_z=0$ and $q_z=0$. In the $k_z=0$ plane or 2D plane,
the wave number vector $\bk=k_x \hat{{\bf x}}+ k_y \hat{{\bf y}}$
and its amplitude $k=\sqrt{k_x^2+k_y^2}$. By choosing the helical
modes $\bh_s(\bk)={\bf N} \btimes \hat{\bk} + is \bnu$ which
satisfy $i\bk\btimes\bh_{\pm} = \pm |\bk|\bh_{\pm}$
(\cite{Waleffe92}), we have \be \bh_s(\bk)=\hat{{\bf z}} + is
\frac{k_y \hat{{\bf x}}-k_x \hat{{\bf y}}}{k}=\hat{{\bf z}} + is
\hat{\bk}^{\perp}. \ee \noindent Here ${\bf N}=\frac{\bk \btimes
\hat{{\bf z}}}{|\bk \btimes \hat{{\bf z}}|}$ and
$\hat{\bk}^{\perp}=(k_y \hat{{\bf x}}-k_x \hat{{\bf y}})/k$ are
unit vectors orthogonal to the unit vector $\hat{\bk}=\bk/k$. The
other two helical modes in the triad are $\bh_s(\bp)=\hat{{\bf z}}
+ is \hat{\bp}^{\perp}$ and $\bh_s(\bq)=\hat{{\bf z}} + is
\hat{\bq}^{\perp}$ , where the unit vectors
$\hat{\bp}^{\perp}=(p_y \hat{{\bf x}}-p_x \hat{{\bf y}}/p$ and
$\hat{\bq}^{\perp}=(q_y \hat{{\bf x}}-q_x \hat{{\bf y}})/q$ are
perpendicular to the unit wave vectors $\hat{\bp}=\bp/p$ and
$\hat{\bq}=\bq/q$, respectively. The velocity projection onto
$\bh_s$ is
\begin{eqnarray}
a_s(k_x,k_y,0,t)& &= \,\left. \frac{\bh_s(k_x,k_y,0) \cdot
\hat{\bu}(k_x,k_y,0)}{2} \right. \cr \, &  & \left. = \,
\frac{1}{2}[ \hat{u}_z(k_x,k_y,0) + i s ( \frac{\hat{u}_x
k_y}{k}-\frac{\hat{u}_y k_x}{k})] \right..
\lb{proj_2D}\end{eqnarray}

\noindent Since \be \hat{\bom}=i\bk \btimes \hat{\bu}=i(k_x
\hat{{\bf x}}+k_y \hat{{\bf y}})\btimes (\hat{u}_x \hat{{\bf
x}}+\hat{u}_y \hat{{\bf y}}+\hat{u}_z \hat{{\bf z}}) = i(k_x
\hat{u}_y - k_y \hat{u}_x) \hat{{\bf z}}=\hat{\omega}_z \hat{{\bf
z}}, \ee \noindent Eq.(\ref{proj_2D}) can be also written as \be
a_{s}(k_x,k_y,0,t) = \frac{1}{2} \hat{u}_z(k_x,k_y,0) +
\frac{s}{2} \frac{\hat{\omega}_z(k_x,k_y,0)}{k}. \ee

\noindent Thus, the vertically-averaged vertical velocity and the
vorticity can be separated from the above equation as
 \be \hat{u}_z=a_+ + a_-,\,\,\ \hat{\omega}_z=k (a_+ - a_-).\lb{2Dvort}\ee

For the slow modes $a_{s_k}(t)=A_{s_k}(t)$, the ``averaged
equation'' Eq.(\ref{averaged}) becomes \be (\partial_t +
\frac{1}{Re} \,\, k^2) a_{s_k}=\frac{1}{4} \sum_{\bk+\bp+\bq=0}
\sum_{s_p,s_q}(s_p p - s_q q)(\bh_{s_p}^* \btimes \bh_{s_q}^*)
\cdot \bh_{s_k}^* a^*_{s_p} a^*_{s_q} \lb{slow-averaged}.\ee
\noindent Since $\hat{\bp}^{\perp} \btimes
\hat{\bq}^{\perp}=\hat{\bp} \btimes \hat{\bq}$,
 \begin{eqnarray}
 (\bh_{s_p}^* \btimes \bh_{s_q}^*) \cdot
\bh_{s_k}^* & & =\,\left.(\hat{{\bf z}} - is \hat{\bp}^{\perp})
\btimes (\hat{{\bf z}} - is \hat{\bq}^{\perp})\cdot (\hat{{\bf z}}
- is \hat{\bk}^{\perp}) \right. \cr \, & & \left.= \,(s_p
\hat{\bp} -s_k \hat{\bk}) \btimes (s_k \hat{\bk}-s_q \hat{\bq})
\cdot \hat{{\bf z}} \right. .
\end{eqnarray}

\noindent Then, Eq.(\ref{slow-averaged}) becomes \be (\partial_t +
\frac{1}{Re} \,\, k^2) a_{ s_k}=\frac{1}{4} \sum_{\bk+\bp+\bq=0}
\sum_{s_p,s_q}(s_p p - s_q q)(s_p \hat{\bp} -s_k \hat{\bk})
\btimes (s_k \hat{\bk}-s_q \hat{\bq}) \cdot \hat{{\bf z}}.\ee
\noindent To construct $\hat{u}_z$ and $\hat{\omega}_z$
(Eq.(\ref{2Dvort})) from the above equation, we have
\begin{eqnarray}
 (\partial_t + \frac{1}{Re} \,\, k^2)(a_{ +_k}- a_{ -_k})  =  \left. \frac{1}{2}
\sum_{\bk+\bp+\bq=0}\frac{p^2-q^2}{k}(\hat{\bp} \btimes \hat{\bq}
\cdot \hat{{\bf z}})[-a_{+_p}^* a_{+_q}^* - a_{-_p}^* a_{-_q}^*
\right. \cr \,
 \left. +  a_{+_p}^* a_{-_q}^* + a_{-_p}^* a_{+_q}^*] \right. ,\lb{2D_1}
 \end{eqnarray}

\noindent and \begin{eqnarray} (\partial_t + \frac{1}{Re} \,\,
k^2) (a_{ +_k} + a_{ -_k}) = \left. \frac{1}{2}
\sum_{\bk+\bp+\bq=0} (\hat{\bp} \btimes \hat{\bq} \cdot \hat{{\bf
z}})[(p-q)(-a_{+_p}^* a_{+_q}^* + a_{-_p}^* a_{-_q}^*) \right. \cr
\,\left. + (p+q)(a_{+_p}^* a_{-_q}^* - a_{-_p}^* a_{+_q}^*)]
\right. .\lb{passive_1}
\end{eqnarray}

Let $b_k=a_{+_k} - a_{-_k}$ and $\theta_k=a_{+_k} + a_{-_k}$, then
$a_{+_k} = (\theta_k + b_k)/2$ and $a_{-_k} = (\theta_k - b_k)/2$.
 Substituting the above relations into Eq.(\ref{2D_1}) and Eq.(\ref{passive_1}),
 we have
\be (\partial_t + \frac{1}{Re} \,\, k^2) b_k =-\frac{1}{2}
\sum_{\bk+\bp+\bq=0}\frac{p^2-q^2}{k}\hat{\bp} \btimes \hat{\bq}
\cdot \hat{{\bf z}} \,\, b_p^* b_q^* \lb{2D_2},\ee

\noindent and \be (\partial_t + \frac{1}{Re} \,\, k^2) \theta_k
=\frac{1}{2} \sum_{\bk+\bp+\bq=0} \frac{\hat{\bp} \btimes
\hat{\bq} \cdot \hat{{\bf z}}}{pq}[q b_p^* \theta_q^* - p b_q^*
\theta_p^*].\lb{passive_2} \ee \noindent Since $\hat{\omega}_z =k
b_k$ from Eq.(\ref{2Dvort}), Eq.(\ref{2D_2}) can be written in
terms of the vorticity as

\be (\partial_t + \frac{1}{Re} \,\, k^2) \hat{\omega}_z
=\frac{1}{2} \sum_{\bk+\bp+\bq=0} \frac{\hat{\bp} \btimes
\hat{\bq} \cdot \hat{{\bf z}}}{pq}[q b_p^* \hat{\omega}_{z_q}^* -
p b_q^* \hat{\omega}_{z_p}^*].\lb{2D_heli-N-S} \ee

\noindent This is exactly the same as the helical formulation of
2D N-S equation which is given by \cite{Waleffe93}.

Also because of $\hat{u}_z(\bk)=\theta_k$ from Eq.(\ref{2Dvort}),
Eq.(\ref{passive_2}) can be written in terms of $\hat{u}_z(\bk)$
as
\be (\partial_t + \frac{1}{Re} \,\, k^2)\hat{u}_{z_k} =
\frac{1}{2} \sum_{\bk+\bp+\bq=0} \frac{(\hat{\bp} \btimes
\hat{\bq} \cdot \hat{{\bf z}})}{pq}(q b^*_p \hat{u}^*_{z_q} - p
b^*_q \hat{u}^*_{z_p}) . \lb{2D_heli-passive} \ee
\noindent This equation has the same structure as the 2D vorticity
equation Eq.(\ref{2D_heli-N-S}). It is known that a 2D passive
scalar and the 2D vorticity obey homologous equations and, in
fact, the above equation is the helical formulation of the 2D
passive scalar equation.

Based on the above discussion, we now summarize the Dynamic
Taylor-Proudman theorem as the following: in three-dimensional
rotating turbulence, when $Ro \rightarrow 0$, ``slow-slow-slow''
triadic interactions yield the 2D-3C Navier-Stokes equations.
Here, 2D-3C refers to two variables ($x$,$y$) and three components
$(\overline{u}^{3D},\overline{v}^{3D},\overline{w}^{3D})$. The
vertically-averaged horizontal velocity
$(\overline{u}^{3D},\overline{v}^{3D})=\frac{1}{H}\int_0^{H}
(u_x(x,y,z),u_y(x,y,z)dz$ satisfies the 2D Navier-Stokes equation
 \be \partial_t \hat \omega_z + \nu k^2 \hat \omega_z=-i k_x \widehat
{\overline{u}^{3D}(x,y) \omega_z(x,y)} - i k_y
\widehat{\overline{v}^{3D}(x,y) \omega_z(x,y)},\lb{2D_N-S_app} \ee
\noindent and the vertically-averaged vertical velocity
$\overline{w}^{3D}=\frac{1}{H}\int_0^{H} u_z(x,y,z)dz$ obeys the
2D passive scalar equation
\be
\partial_t \widehat{\overline{w}^{3D}}(\bk) +
\nu k^2 \widehat{\overline{w}^{3D}}(\bk) = -i k_x \widehat
{\overline{u}^{3D}(x,y) \overline{w}^{3D}(x,y)} - i k_y \widehat
{\overline{v}^{3D}(x,y) \overline{w}^{3D}(x,y)}.
\lb{2D_passive_app}\ee
\noindent Here, $H$ is the vertical height of the domain and the
three averaged velocity components in the Fourier space are as
follows:
\be \widehat{ \overline{u}^{3D}}(\bk)= i \frac{k_y}{k^2_x+k^2_y}
\hat \omega_z(\bk)\,\,,\widehat{\overline{v}^{3D}}(\bk)= -i
\frac{k_x}{k^2_x+k^2_y} \hat
\omega_z(\bk),\,\,\widehat{\overline{w}^{3D}}(\bk)=\hat{u}_z(\bk).\ee

\end{document}